\def\spose#1{\hbox to 0pt{#1\hss}}
\def\approxlt{\mathrel{\spose{\lower 3pt\hbox{$\sim$}}
	\raise 2.0pt\hbox{$$<$$}}}
\def\approxgt{\mathrel{\spose{\lower 3pt\hbox{$\sim$}}
	\raise 2.0pt\hbox{$>$}}}

\def\multleft#1{\hbox to size{\vbox {\halign {\lft{##}\cr #1}}\hfill}\par}
\def\multright#1{\hbox to size{\vbox {\halign {\rt{##}\cr #1}}\hfill}\par}

\def\today{\ifcase\month\or January\or February\or March\or April\or May\or
      June\or July\or August\or September\or October\or November\or December\fi
      \space\number\day, \number\year}
\def\$<${\thinspace}
\def\s{\hbox{\phantom{5}}}	

\def\boxit#1{\vbox{\hrule\hbox{\vrule\kern3pt\vbox{\kern3pt
          #1 \kern3pt}\kern3pt\vrule}\hrule}}

\def\cm{{\rm\thinspace cm}}

\def\erg{{\rm\thinspace erg}}
\def\eV{{\rm\thinspace eV}}

\def\K{{\rm\thinspace K}}
\def\keV{{\rm\thinspace keV}}
\def\km{{\rm\thinspace km}}
\def\kpc{{\rm\thinspace kpc}}
\def\kpcsq{{\rm\thinspace kpc$^{2}$}}
\def\Lsun{\hbox{$\rm\thinspace L_{\odot}$}}

\def\Mpc{{\rm\thinspace Mpc}}

\def\pc{{\rm\thinspace pc}}

\def\s{{\rm\thinspace s}}

\def\pcmcu{\hbox{$\cm^{-3}\,$}}

\def\ergpcmsqps{\hbox{$\erg\cm^{-2}\s^{-1}\,$}}

\def\ergps{\hbox{$\erg\s^{-1}\,$}}

\def\kmps{\hbox{$\km\s^{-1}\,$}}

\def\pcm{\hbox{$\cm^{-3}\,$}}

\def\ps{\hbox{$\s^{-1}\,$}}
\def\psqcm{\hbox{$\cm^{-2}\,$}}

\def\kmpspMpc{\hbox{$\kmps\Mpc^{-1}$}}

\documentstyle[psfig]{mn}
\begin{document}


\title{Mapping the gas kinematics and ionization structure of four ultraluminous IRAS galaxies}

\author[]
{\parbox[]{6.in} {R.J.~Wilman, C.S.~Crawford \& R.G.~Abraham\\ 
\footnotesize
Institute of Astronomy, Madingley Road, Cambridge CB3 0HA \\ }}

\maketitle

\begin{abstract}
We present a study of the morphology, kinematics and ionization structure of the extended emission-line regions in four intermediate redshift 
($0.118<z<0.181$) ultraluminous infrared galaxies, derived from ARGUS two-dimensional fibre spectroscopy. 

The gas kinematics in the hyperluminous system IRAS F20460+1925 lack coherent structure, with a $\rm{{\em FWHM}}>1000$\kmps~within 1 arcsec of the nucleus, suggesting that any merger is well-advanced. Emission-line intensity ratios point to AGN photoionization for the excitation of this gas at the systemic velocity. An isolated blob $\sim8$\kpc~from the nucleus with a much smaller velocity dispersion may lie in a structure similar to the photoionization cones seen in lower-luminosity objects. A second, spatially-unresolved, narrow line component is also present on nucleus, blueshifted by $\simeq 990$\kmps from the systemic and plausibly powered by photoionizing shocks.

IRAS F23060+0505 has more ordered kinematics, with a region of increased {\em FWHM} coincident with the blue half of a dipolar velocity field. The systemic velocity rotation curve is asymmetric in appearance, due either to the on-going merger or to nuclear dust-obscuration. From a higher-resolution ISIS spectrum, we attribute the blue asymmetry in the narrow line profiles to a spatially-resolved nuclear outflow. Emission-line intensity ratios suggest shock+precursor ionization for the systemic component, consistent with the X-ray view of a heavily obscured AGN. 

The lower luminosity objects IRAS F01217+0122 and F01003-2238 complete the sample. The former has a featureless velocity field with a high {\em FWHM}, a high-ionization AGN spectrum and a $\sim 1$~Gyr old starburst continuum. IRAS F01003-2238 has a dipolar velocity field and an HII-region emission line spectrum with a strong blue continuum. After correction for intrinsic extinction, the latter can be reproduced with $\sim 10^{7}$ O5 stars, sufficient to power the bolometric luminosity of the entire galaxy.

We accommodate this diversity within the merger-induced evolutionary scenario for ultraluminous infrared galaxies: the merger status is assessed from the kinematics in a way which is consistent with morphological and colour information on the galaxies, or with the inferred ages of the young stellar populations and the dominance of the AGN.  
\end{abstract}


\section{INTRODUCTION}
One of the most important results to stem from the IRAS mission of the early 1980s was the discovery of a population of galaxies which emit most of their bolometric luminosity at mid- to far-infrared wavelengths (see Sanders \& Mirabel 1996 for a review). Whilst there is little doubt that this is thermal re-radiation from heated dust grains, the nature of the underlying power source remains uncertain -- for some individual objects and for the class of ultraluminous infrared galaxies (ULIRGs, with $L_{\rm{IR}}>10^{12}$\Lsun) as a whole. The resolution of this problem has significant implications for the wider context of extragalactic astronomy. For example, some of the most luminous ULIRGs and some hyperluminous objects ($L_{\rm{IR}}>10^{13}$\Lsun) exhibit high-ionization Seyfert-2-like optical spectra [eg. IRAS F15307+3252, Hines et al.~(1995); IRAS P09104+4109, Hines \& Wills~(1993)], leading to suggestions that they constitute the long-sought `type-2' quasars -- active galactic nuclei (AGN) with Seyfert 2-like optical spectra but at QSO luminosities. The results of mid-infrared ISO spectroscopic surveys (Genzel et al. and Lutz et al.~1998), however, suggest that more than 80 per cent of ULIRGs are predominantly powered by recently-formed massive stars, with the AGN-powered fraction increasing to around half above an infrared luminosity of $2 \times 10^{12} \Lsun$.   

For the past decade, the formation and evolution of ULIRGs have been discussed in the context of the paradigm of Sanders et al.~(1988a), in which they constitute an early phase in the merger-induced formation of optically-selected quasars. Following the merger of two gas-rich spiral galaxies, tidal torques drive material into the merger nuclei, leading to prodigious star formation and the fuelling of supermassive black holes. The circumnuclear regions are subsequently swept clear of obscuring dust by radiation pressure, exposing the quasar nucleus. Whilst this model was in part suggested by the frequency of occurrence of ULIRGs in interacting or merging systems (Sanders et al.~1988b), since corroborated in the study of larger samples (eg. Clements et al.~1996), it is undoubtedly not the whole story. Indeed, from their ISO spectroscopy, Genzel et al.~(1998) found no obvious trend for the AGN component to dominate in the more advanced mergers of their sample, implying that merger-phase alone does not determine the relative dominance of an AGN component. Similarly, the Fabry-Perot H$\alpha$ velocity mapping of four ULIRGs by Mihos \& Bothun~(1998) showed that ultraluminous activity is not confined to late stage mergers, suggesting that details such as the internal structure or gas content of the merging galaxies can strongly influence their evolution. 

In a study of the extended emission line region (EELR)
 of the ULIRG IRAS P09104+4109 with the ARGUS integral field spectrograph on the Canada France Hawaii Telescope (CFHT), Crawford \& Vanderriest (1996) found that the gas was energised by an obscured quasar continuum and suggested that a 1200 $\kmps$ nuclear outflow was responsible for displacing X-ray-emitting gas from the centre of the surrounding cooling-flow cluster. These findings prompted us to use ARGUS to study a small sample of ULIRGs, in order to discover whether the properties of IRAS P09104+4109 are typical of such luminous objects or due to its unusual environment, and whether any trends with increasing luminosity can be identified.  

The two principal targets in our new sample are the famous QSO-like systems IRAS F20460+1925 and F23060+0505, at redshifts of 0.181 and 0.174 respectively; they were chosen partly for their large luminosities (falling just either side of the conventionally-adopted ultra/hyperluminous divide) and also because they have been studied extensively at other wavelengths, with evidence from both X-ray studies and near-IR broad emission lines for obscured active nuclei. Included for comparison are two lower-luminosity, lower-redshift objects, IRAS F01217+0122 and F01003-2238, the latter having Wolf-Rayet features in its spectrum, suggestive of a decaying starburst (Armus, Heckman \& Miley 1988). The primary aim of the present area-spectroscopic study is to understand the observed kinematics and morphology of these EELRs within the context of plausible dynamical models for their origin in, for example, galaxy mergers. Secondly, we seek to assess the viability of various ionization mechanisms, including power-law photoionization, fast shocks and their precursor HII regions, and young massive stars. 

For each of the two most luminous objects, we discuss the views provided by the ARGUS data on the gas kinematics and its ionization structure. We also present higher resolution long-slit spectra of these two objects taken with the William Herschel Telescope (WHT), and comment upon multicolour images from the Jacobus-Kapteyn Telescope (JKT). In obtaining the former, we sought both to confirm the ARGUS findings and to overcome some of their inherent limitations arising from their modest spectral resolution and decreased signal-to-noise towards the blue part of the spectrum. Following a less detailed presentation of the ARGUS data for the other two objects, we conclude with a discussion of the comparative properties of the sample as a whole. For convenience, we summarise in Table 1 the lay-out of the results section.

\begin{table*}
\begin{center}
\caption{Lay-out of the results sections of this paper}
\begin{tabular}{||lllll||} \hline
 & F01003-2238 & F01217+0122 & F20460+1925 & F23060+0505 \\ \hline 
ARGUS: &      6.2        &    6.1          &  3      & 4    \\
gas kinematics &   6.2   &   6.1      &    3.1.1   & 4.1.1    \\
ionization structure & 6.2 &  6.1      &  3.1.2     &  4.1.2  \\ \hline
ISIS:               &  --     &   --      & 3.2 & 4.2    \\ \hline
JKT:         &  --   &    --    &  5   & 5 \\ \hline
\end{tabular}
\end{center}
\end{table*}

The cosmological parameters $H_{\rm{0}}=50\kmpspMpc$ and $q_{\rm{0}}=0.5$ have been adopted throughout.

\section{OBSERVATIONS, DATA REDUCTION \& ANALYSIS}

The details of all the observations are presented in Table 2 and discussed separately below.

\begin{table*}
\begin{center}

\caption{Summary of the observations}
\begin{tabular}{||lllll||} \hline
 & F01003-2238 & F01217+0122 & F20460+1925 & F23060+0505\\ \hline 
z & 0.118 & 0.137 & 0.181 & 0.174 \\ 
log($L_{\rm{FIR,Bol}}/\Lsun$) & 12.2$^{\dagger}$ & 12.4$^{\dagger}$ & 13.2$^{\ddagger}$ & 12.9$^{\star}$ \\ \hline \hline
\multicolumn{5}{c}{ARGUS integral field spectroscopy on the CFHT} \\ 
\multicolumn{5}{c}{1996 August 14--15} \\ 
\multicolumn{5}{c}{Spectral resolution: 10.3~\AA~{\em FWHM}} \\ 
\multicolumn{5}{c}{Wavelength coverage: 4000--10000~\AA} \\ \hline
Seeing: arcsec & 1.5--2.0 & 0.7--1.5 & 0.5 & 0.5--0.7 \\ 
Integration time: s & 3000 & 5400 & 10800 & 9000 \\ \hline \hline
\multicolumn{5}{c}{ISIS longslit spectroscopy on the WHT} \\ 
\multicolumn{5}{c}{1998 September 21} \\ 
\multicolumn{5}{c}{Spectral resolution: 3.1~\AA~{\em FWHM}} \\ 
\multicolumn{5}{c}{Wavelength coverage: 3650--6000~\AA} \\ \hline
Seeing: (arcsec) & -- & -- & $\simeq1$ & $\simeq1$ \\ 
Integration time: (s) & -- & -- & 3600 & 3600 \\ \hline \hline
\multicolumn{5}{c}{JKT multi-band imaging} \\ 
\multicolumn{5}{c}{1998 August 20--21} \\ \hline
Seeing: (arcsec) & -- & -- & $\simeq1$ & $\simeq1$ \\ 
B-band Integration time: (s) & -- & -- & 2000 & 2000 \\  
R-band Integration time: (s) & -- & -- & 1200 & 1200 \\  
I-band Integration time: (s) & -- & -- & 2000 & 1800 \\ \hline \hline
\end{tabular}
\end{center}
$\dagger$ $L_{\rm{FIR}}=L(40-500 \mu m)$, computed from IRAS fluxes using the formula in Sanders \& Mirabel~(1996).\\
$\ddagger$ $L_{\rm{Bol}}$ computed by Frogel et al.~(1989)\\
$\star$ $L(0.5-120 \mu m)$, given by Hill et al.~(1991)\\
\end{table*}

\subsection{ARGUS integral field spectroscopy}

Observations were made with the ARGUS integral field spectrograph on the CFHT during the nights of 1996 August 14 and 15. The instrument consists of 
a 12.8 $\times$ 7.8 $\rm{arcsec}^2$ aperture in the focal plane of the 
telescope, within which 0.4 arcsec diameter optical fibres are arranged into 27 rows. At the entrance to the multi-object spectrograph the fibre bundle is drawn out to mimic a long slit arrangement with adjacent rows being separated by blank fibres. The O300 grism and STIS 2 CCD were employed, covering the range 4000-10000\AA~at a dispersion of 5\AA~per pixel and a resolution of approximately 10.3\AA~{\em FWHM}. A more complete description of the instrumental set-up and performance is provided by Crawford and Vanderriest (1997).

The principal targets, IRAS F20460+1925 and F23060+0505, were observed for 
total exposures of 10800 and 9000 seconds, respectively. The seeing during this period was typically 0.5-0.7 arcsec {\em FWHM}, but it deteriorated three hours into the second night to $\sim 1.5-2$ arcsec {\em FWHM} when the lower luminosity objects, IRAS F01003-2238 and F01217+0122, were observed (for 3000 and 5400 seconds, respectively).  

The reduction of the ARGUS data was performed within IRAF. Cosmic rays were removed by co-adding the separate images taken of each field. Thereafter, a two-dimensional transformation was applied to the CCD frames in order to correct for a form of optical distortion in which the blank spacer fibres were seen to bow outwards towards the left- and right-hand edges of the frame. Each fibre spectrum then projected onto two rows of the chip across the wavelength range. Wavelength calibration was performed by applying a wavelength-pixel number transformation derived from He/Ne, He/Ar, He/Ne/Ar and Ar arc spectra. Bias subtraction was problematic, as the level varied throughout each night and with spatial location on the chip. In consequence, a separate bias frame was synthesised for each data frame as follows: all of the good bias frames with similar (x,y) structure were combined to produce a normalised bias frame, which was then multiplied by the y-variation of the x-averaged bias level for the data frame under consideration (as estimated from the blue region of the chip where the quantum efficiency of the system is low). Flat-fielding was not possible due to features in the flat-field exposures (possibly induced by local heating of the CCD) which rendered impossible their bias subtraction according to the above procedure. Compensation for spectrally-averaged fibre-to-fibre sensitivity variations was made at a later stage of the analysis. This was done by using an arc-lamp exposure to select those fibres with a wavelength-integrated flux reduced by more than $\sim 10$ per cent with respect to adjacent fibres. The derived transmission correction factors for such fibres were later used to scale their observed emission lines fluxes. This local method does not correct for global variations in the sensitivity of the system across the aperture, as they cannot be distinguished from any spatial gradient in the arc light illumination.
 
Although spectra were taken of six flux standard stars, it was subsequently found
 that only two have fluxes tabulated up to $10000$\AA~(the rest terminate at 
$8400$\AA) and, furthermore, that one of these is contaminated by an unphysical feature due to a second order of interference. We thus had only one standard star covering the spectral range of interest. At wavelengths 
less than $8400$\AA~a flux uncertainty of $\leq 25$ per cent is estimated, 
based upon the scatter between the flux sensitivity curves in this region.
 Corrections were made for atmospheric extinction and Galactic reddening, 
assuming for the latter the $N_{\rm{H}}:E(B-V)$ conversion of Bohlin, Savage 
$\&$ Drake (1978) and an R value of 3.2 (line-of-sight Galactic hydrogen 
densities were obtained from Stark et al. 1992). The CCD frames were split 
into individual fibre spectra using an ARGUS-dedicated IRAF package. Sky subtraction used an average sky spectrum for each row of the hexagon, constructed from the mean of four or five fibre spectra in each of that row and the adjacent two. Care was taken to avoid fibres in which the effects of aperture vignetting were not negligible. 

Comparison of the emission-line fluxes derived from the ARGUS data with previously-published values indicates that our fluxes are systematically larger, by at least 50 per cent. This may be partly due to the fact that previous long-slit spectra did not admit all of the extended emission. Comparison of the [OIII] and [OII] fluxes with those of Hough et al.~(1991) for IRAS F23060+0505 and with the ISIS data (section~2.2) confirms the reliability of the relative flux calibration of the ARGUS data blueward of redshifted [OIII]$\lambda5007$ at $\sim 5900$\AA. The ARGUS flux calibration does not vary across the aperture, and errors in its absolute level do not significantly affect the scientific conclusions of this paper. The latter depend mainly upon the kinematic decomposition of individual line profiles and upon flux ratios of lines either close in wavelength (eg. [OIII]/H$\beta$) or from spectral regions over which the relative flux calibration is reliable (eg. [OIII]/[OII]).   

\subsection{ISIS spectroscopy}
Higher-resolution long-slit spectra of IRAS F20460+1925 and F23060+0505 were obtained with the blue arm of the ISIS spectrograph on the 4.2m WHT during the night of 1998 September 21. A one hour integration was performed for each object with the R300B grating in place. In conjunction with the EEV12 CCD this yielded a wavelength coverage of 3650--6500\AA~at a spatial scale of 0.2 arcsec per pixel and a spectral resolution of 3.1\AA~{\em FWHM}. The width of the slit was matched to the seeing ($\simeq1$~arcsec) and oriented at a position angle chosen to pass through features of potential interest (described later).   

The data reduction was performed within IRAF. Following bias subtraction, the data frame was rotated by $\simeq 0.3$~degrees to bring the slit and dispersion axes into more exact alignment with those of the chip. The spectra were subsequently wavelength-calibrated using a CuAr arclamp exposure, flux-calibrated using observations of the standard star HD192281, corrected for atmospheric extinction and Galactic reddening, and sky-subtracted. Since no structure could be discerned in the flat field frames, the data were not flat-fielded. 

\subsection{JKT multicolour imaging}

Using the 1m Jacobus-Kapteyn Telescope (JKT), we obtained images of IRAS F20460+1925 and F23060+0505 in the B, R and I bands. The observations were carried out on the nights 1998 August 20--21, with integration times of between 1200 and 2000 seconds per object, per filter, in seeing conditions of approximately 1 arcsec {\em FWHM}. The Tek4 CCD was employed, covering a 5.6 square arcmin field of view on a scale of 0.33 arcsec per pixel. 

The preliminary reduction of these data, namely bias and sky-subtraction, was performed within IRAF, but they were not considered to be of sufficient quality to merit flux-calibration and detailed quantitative analysis. 

\subsection{Emission-line fitting: some general remarks}
The most prominent emission line complexes in the ARGUS spectra of all five objects
 are those of [OIII]$\lambda4959,5007$+H$\beta$ and H$\alpha$+[NII]$\lambda6548,6584$. The 
[OII]$\lambda3727$ doublet is detected only in the nuclear fibres as the signal-to-noise ratio deteriorates significantly towards the blue end of the spectrum. Due to deblending uncertainties in the [NII]+H$\alpha$ complex of these relatively low resolution spectra, kinematic information was extracted only from fits to the [OIII]+H$\beta$ profile. Fitting was performed using the QDP package (Tennant 1991). 

In the first instance, the narrow lines in a complex were fitted with single component gaussian profiles having the same velocity width and redshift. The [OIII]$\lambda 5007/4959$ flux ratio was set to the theoretical value of 3:1 and the continuum was modelled as a constant. The fitting of such a profile facilitated the rapid identification of the gross kinematical features, providing some indication of those aspects worthy of more diligent fitting with more complex models.
 Many of the emission lines were found to be asymmetric in appearance and thus
 poorly described by a single gaussian. In most such cases the total line profiles could be
 satisfactorily modelled by a function consisting of two [OIII]+H$\beta$ 
complexes of the above form with a fixed velocity offset between them, and
 with the [OIII] and H$\beta$ fluxes of the second component expressed as fractions of those in the first. The velocity widths of all lines in the two 
components were, however, constrained to be equal. Note that in fitting a double component profile we do not wish to imply that there are precisely two physically-distinct components to the emission. Asymmetric and multi-component line profiles are a common feature of the narrow line spectra of Seyfert galaxies and a discussion of the complexities associated with their physical interpretation is given by, for example, Veilleux (1991).

The decomposition of the H$\alpha$+[NII] complex was not always possible, because the presence of multiple velocity components, and in some cases broad H$\alpha$, frequently rendered the entire feature unresolved. Where it could be modelled with reasonable confidence, multiple velocity component profiles were employed, constructed in exact analogy to those described above for the [OIII]+H$\beta$ complex. Specifically, the lines within both velocity components were assumed to have identical velocity widths and the [NII]$\lambda$6584/6548 flux ratio was set to the theoretical value of 3:1. Any broad H$\alpha$ was assumed to be at the redshift of the systemic narrow line component. The [OII]$\lambda$3727 doublet was modelled as a single gaussian, which is a satisfactory approximation at this resolution.   

In order to understand how systematic effects might affect the interpretation of the parameter values extracted from fits with simple models, we performed some simulations. Synthetic, noise-free, profiles for the [OIII]+H$\beta$ complex were generated and fitted with the above double component model. The findings are alluded to at various points in the text, and discussed in detail in Appendix A. The higher resolution of the ISIS spectra allowed us to fit the [OIII]+H$\beta$ complex with a model in which the velocity widths of the two components were not constrained to be the same, thus overcoming these systematic effects. The ISIS data also have significantly better wavelength coverage and signal-to-noise blueward of H$\beta$ than the ARGUS data, thus permitting the analysis of emission lines such as those of [OII]$\lambda 3727$, [NeIII], [NeV], HeII and higher order Balmer lines. Where two (as opposed to just a single) gaussian velocity components could justifiably be fitted to such profiles, their redshifts and widths were fixed at the values derived from the [OIII]+H$\beta$ complex.

The position-dependent instrumental resolution was subtracted in quadrature from the fitted emission-line widths, with the former determined from fits to groups of arc lines at wavelengths similar to the galactic emission-lines under consideration. All {\em FWHM} quoted in this paper have thus been corrected for this instrumental effect. 

There are two additional factors which serve to complicate the profile fitting and its subsequent interpretation. The first is that 
projection effects result in the observed profile at any point in the sky 
being due to emission from all depths along the line of sight. The second is a consequence of the finite size of 
the point spread function (PSF) of the telescope optics convolved with the atmospheric seeing, as sampled by the fibre arrangement: the light from a point source is collected by several fibres. As a result, the profile observed in a given fibre is contaminated by light from points on the sky centred on neighbouring fibres. A proper understanding of this effect is necessary when the observed object consists of a bright point source enveloped by low surface brightness diffuse emission, as is the case for several of the ULIRGs in our sample. The standard star exposures taken at frequent intervals enable the structure of the PSF to be monitored: its {\em FWHM} was typically found to be larger along the rows of fibres than in the perpendicular direction (see Crawford \& Vanderriest 1996). 

All errors quoted in this paper are $\Delta \chi^{2}=2.7$ intervals for one parameter of interest (corresponding to 90 per cent confidence), assuming that the original fit was good. They are fitting errors only and do not incorporate any systematic effects from the data reduction. 

\section{RESULTS FOR IRAS F20460+1925}

IRAS F20460+1925 was selected from the IRAS catalogue by Frogel et al. (1989) as part of a programme to study sources with flat far-infrared spectral energy distributions. They classified it as a Seyfert 2 galaxy at z=0.181 with exceptionally broad permitted and forbidden lines ({\em FWHM} between 1700 and 2100$\kmps$). Hines (1991) and Veilleux et al. (1997) detected broad Pa$\alpha$ with {\em FWHM} $\sim 3000\kmps$ and deduced intrinsic extinctions to the hidden BLR of $A_{\rm{V}} > 3.6$ and $> 6.4$~mag, respectively (cf. the $A_{\rm{V}}=2.4$ mag to the NLR found by Frogel et al.). These values may not represent that along the line of sight to the BLR if a substantial portion of the BLR spectrum is contributed by scattered light, as is suggested by Kay \& Miller's (1989) detection of broad H$\beta$ in polarised light. 

Ogasaka et al. (1997) reached a similar conclusion from their 2-10$\keV$ ASCA spectrum, which they modelled as a power-law with photon index $\Gamma \simeq 2$ along with iron K emission and excess photoelectric absorption of $N_{\rm{H}} \simeq 2.6 \times 10^{22} \psqcm$. They deduced, however, an absorption-corrected value of $L_{\rm{X}}/L_{\rm{Bol}}$ an order of magnitude below that of a typical AGN, perhaps because the direct emission is completely blocked by an obscuring torus; that 
even the scattered X-rays are then absorbed indicates that the nucleus of this 
galaxy is indeed heavily obscured. Alternatively, it could be that a starburst contributes significantly to the bolometric luminosity, but their data neither support nor refute the existence of an additional soft thermal component. 
 
Young et al. (1996a) included the object in their sample of ``warm'' IRAS 
galaxies with Seyfert-2-like optical spectra for which they presented polarimetry and modelling. They found multicomponent [OIII]+H$\beta$ lines and an [NII]+H$\alpha$ complex which could be approximated by double component narrow 
lines along with broad H$\alpha$. Based on measurements of the broad and
 polarised H$\alpha$ fluxes, they deduced an intrinsic polarisation of 12 
per cent and an extinction of 1.2 mag in the scattered light. Along the line
 of sight to the near-IR-emitting regions they calculated an $A_{\rm{V}}$ of
 14 mag, consistent with the lower limit on the extinction to the BLR deduced by Hines (as the BLR is expected to lie within the near-IR-emitting region).

\subsection{The ARGUS data}
\subsubsection{Kinematics and spatial distribution of the line-emitting gas}
\begin{figure}
\psfig{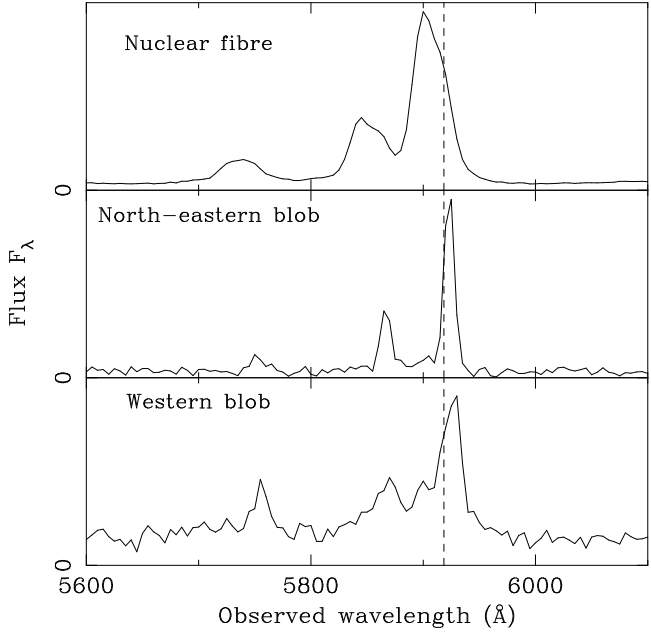}
\caption{\normalsize The [OIII]+H$\beta$ complex in the nuclear fibre, and the kinematically distinct north-eastern and western blobs in IRAS F20460+1925. The position of the nuclear fibre systemic velocity is indicated by the dashed line.}
\label{fig:F20argus}
\end{figure}

An examination of the CCD frames for this object prior to their decomposition
 into fibre spectra clearly showed the double component nature of the [OIII]+H$\beta$ complex in and around the nucleus. A localised region of much narrower, single component emission, spatially isolated from the nucleus and redshifted relative to it, could also be discerned and will henceforth be referred to as the {\em north-eastern blob}. Fig.~\ref{fig:F20argus} shows the [OIII]+H$\beta$ complex of the nuclear fibre, the north-eastern blob and an additional kinematically distinct area to be referred to later as the {\em western blob}.

\begin{figure}
\psfig{figure=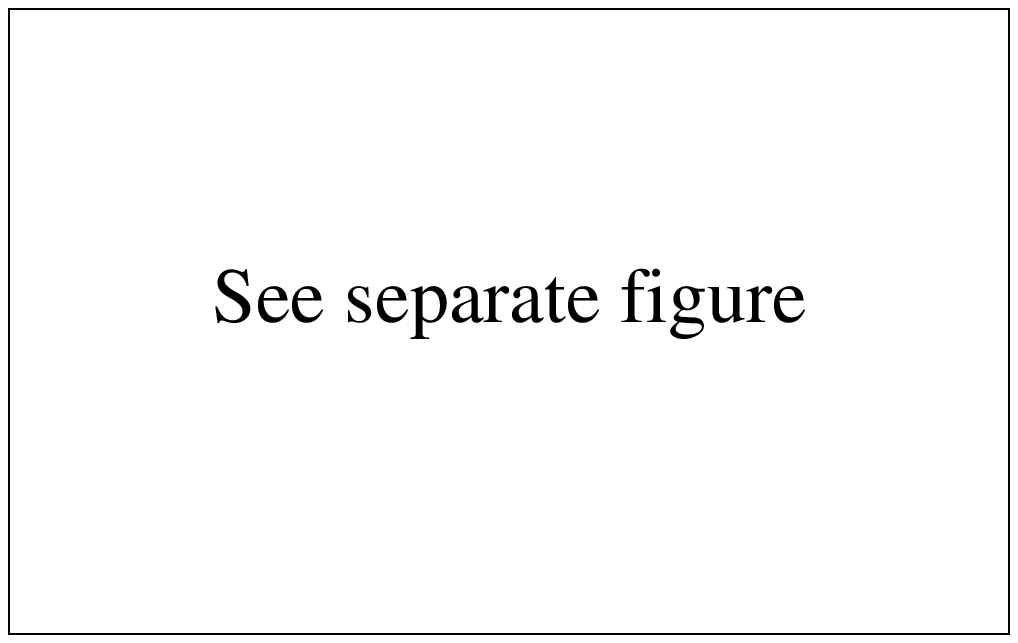,width=0.48\textwidth,angle=270}
\caption{\normalsize Reconstructed images of IRAS F20460+1925 in various emission lines, as described in the text. The square root of the integrated line flux has been plotted on a common scale (in units of $10^{-8}$erg$^{0.5}$cm$^{-1}$s$^{-0.5}$) in order to enhance the dynamic range; the north-eastern blob is highlighted with diamonds. Discs are drawn only for those fibres with a significant detection. The actual flux ranges spanned by the fibres shown are (in units of $10^{-15}$\ergpcmsqps): 0.021--8.53 (for systemic [OIII]), 0.063--14.0 (for blueshifted [OIII]), 0.16--8.26 (for [OII]), and 0.063--7.48 (for [NII]+H$\alpha$). All maps peak on the same fibre, the nucleus, marked with a cross. North is at the top and east to the left in this figure.}
\label{fig:F20inten}
\end{figure}

When the spectra were fitted with the double component [OIII]+H$\beta$ model, it was found that two components were present out to a radius of $\sim 2$~arcsec from the fibre of peak intensity (the nucleus) [1 arcsec $\equiv 4\kpc$ at z=0.181]. In the nucleus itself, the blueshifted component is offset from the systemic by $\simeq -975 \kmps$ with relative intensities in [OIII] and H$\beta$ of $\simeq$ 1.6 and 0.9 respectively, implying that it is the more highly ionized of the two components. Fig.~\ref{fig:F20inten} shows reconstructed images of the object in several (continuum-subtracted) emission lines. [OII]$\lambda3727$ could be detected only within a limited region and reliably decomposed into two velocity components over a yet smaller area, so the flux in the [OII] map is that extracted from a fit to a single gaussian. A rigorous fitting of the [NII]+H$\alpha$ complex could not be performed because of the difficulty of deblending the two (roughly equal-intensity) components to each line and the broad H$\alpha$; we thus map the flux in a (continuum-subtracted) cut between 7730 and 7780~\AA, wide enough to include the broad H$\alpha$. The fluxes in all of the lines peak on the nuclear fibre (marked by a cross) for both the systemic and blueshifted components and exhibit the same, essentially circular, morphology. The total luminosity of the system in the [OIII]$\lambda 5007$ line is $2.56 \times 10^{43} \ergps$, 54 per cent of which comes from the blueshifted component. 

Although a blueshifted component to the [OIII] emission is detected 
in fibres up to 2 arcsec from the nucleus, this may not necessarily indicate an outflow with a radius of 8\kpc. Caution is necessary because the morphology of the second component superficially resembles that of the telescope PSF and thus might in fact be unresolved. To investigate this, a PSF for the wavelength interval 5500-6000\AA~was constructed from the observation of the flux standard BD404032 taken immediately after the first of the IRAS F20460+1925 fields (and thus under comparable seeing). The intensity of the blueshifted component (relative to that in the nuclear fibre) was measured in eight fibres at various distances from the nucleus and compared with that expected from a point source. Within the sampling uncertainties arising from the fact that the standard star PSF appears to peak {\em between} fibres whereas all the intensity maps in Fig.~\ref{fig:F20inten} peak sharply {\em on} a single fibre (to say nothing of the profile fitting uncertainties), the second component appears unresolved. A further argument in support of this conclusion is that the line-of-sight velocity of an outflow with any reasonable geometry would exhibit spatial structure (eg. decreasing to zero at the edges for a spherical bubble), whereas the observed wavelength of the blue component is constant (to within the profile-fitting uncertainties) in all fibres where it is seen.     

\begin{figure}
\psfig{figure=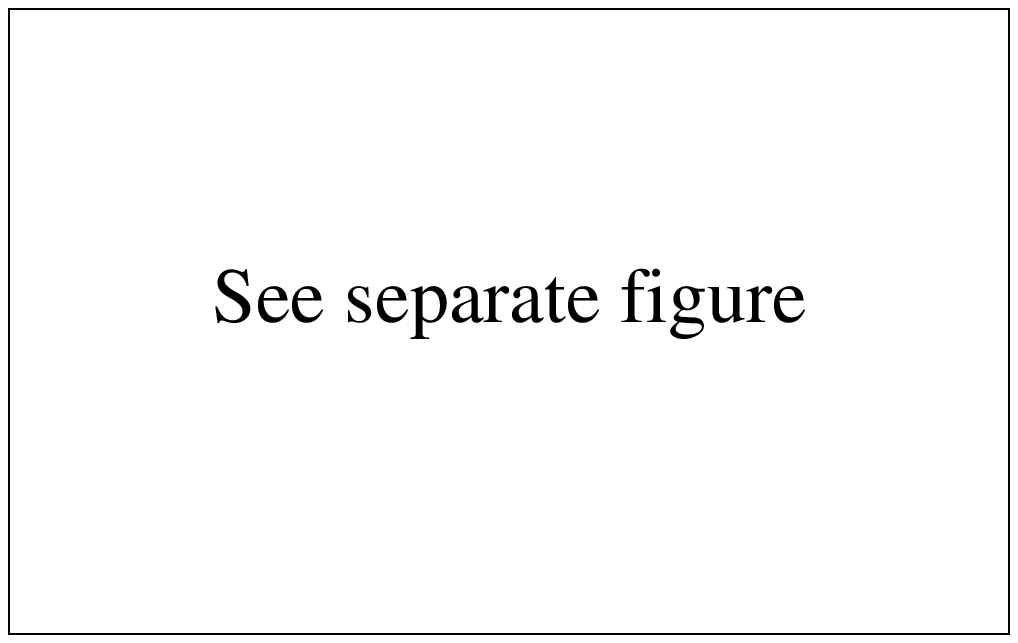,width=0.48\textwidth,angle=270}
\caption{\normalsize The upper map shows the velocity width {\em FWHM} common to the systemic and blueshifted components in a two-component fit to the [OIII]+H$\beta$ complex of IRAS F20460+1925, corrected for the position-dependent instrumental resolution. Those fibres marked with asterisks have an unresolved linewidth (see text). The lower map shows the radial velocity of the systemic component of [OIII]+H$\beta$ emission relative to the nuclear fibre (marked with a cross). North is at the top and east to the left in this figure, and the kinematically distinct north-eastern and western blobs are indicated.}
\label{fig:F20kin}
\end{figure}

Fig.~\ref{fig:F20kin} maps the velocity and {\em FWHM} of the systemic [OIII]+H$\beta$ component. An asterisk on a fibre indicates that the fitted width is consistent with the profile being spectrally unresolved and for such fibres, the plotted {\em FWHM} is taken as the average of that in neighbouring fibres. From this figure, it can be seen that the velocity of the systemic component is substantially constant across the nebula, apart from two kinematically distinct regions. The north-eastern blob is centred at a projected radius of 8$\kpc$, with single component emission redshifted by $\simeq 240 \kmps$ relative to the nucleus, and the western blob, an approximately elliptical region to the west of the nucleus, is at a relative velocity of $\simeq 400 \kmps$. Although the velocity separation of the two components is poorly-constrained in the western blob, it is several hundred $\kmps$ larger than it is in the nucleus. As the latter figure is comparable to the difference in the systemic velocities of the nucleus and the blob itself, this is consistent with the blueshifted component being unresolved. 

It can be seen in Fig.~\ref{fig:F20kin} that, with the exception of fibres in the north-eastern blob wherein just a single narrow (and in places unresolved) component of emission is seen, the lines have a {\em FWHM} of several hundred \kmps, exceeding $1000\kmps$ within $\simeq1$~arcsec of the nucleus. Such large velocity dispersions suggest that the gas is not in a disk-like structure and that it has significant motions in all three spatial dimensions. This situation is to be contrasted with the findings of Unger et al. (1987) who, in their study of the extended narrow line regions of Seyfert 2 galaxies, deduced that the kinematics of the gas within the ionization cones of such objects were consistent with normal galactic rotation. Typical velocity dispersions for neutral gas in the discs of normal spiral galaxies are 7-10 \kmps (van der Kruit \& Shostak~1984). For IRAS F20460+1925 it is remarkable that, on kiloparsec scales, the {\em FWHM} exceeds the upper end of the range ($300-500\kmps$) found for classical narrow line regions in AGN, which are typically $10-1000 \pc$ in size. 
 
There is a caveat to our interpretation of the fitted velocity field. Due to the relatively low resolution of the ARGUS spectrograph, the systemic and blueshifted components were constrained to have the same width. It was, however, discovered during emission-line fitting simulations (see Appendix A) that the parameters extracted from such fits systematically mis-represent the true properties of the two components (especially their relative intensities and velocity separation) {\em if} their widths are in fact markedly different. This qualifies our deduction that the second component is spatially unresolved, and reduces the physical significance which should be attached to the magnitude of its measured velocity offset from the systemic. That the blueshifted component appears to be spatially unresolved implies that its profile results from integration over an entire velocity field and thus suggests that it may be much broader than the systemic. The resolution of these potential sources of ambiguity was one of the motivations behind our obtaining the higher signal-to-noise ISIS spectrum of this object, which is discussed in section 3.2. 

\subsubsection{Ionization structure}

\begin{figure}
\psfig{figure=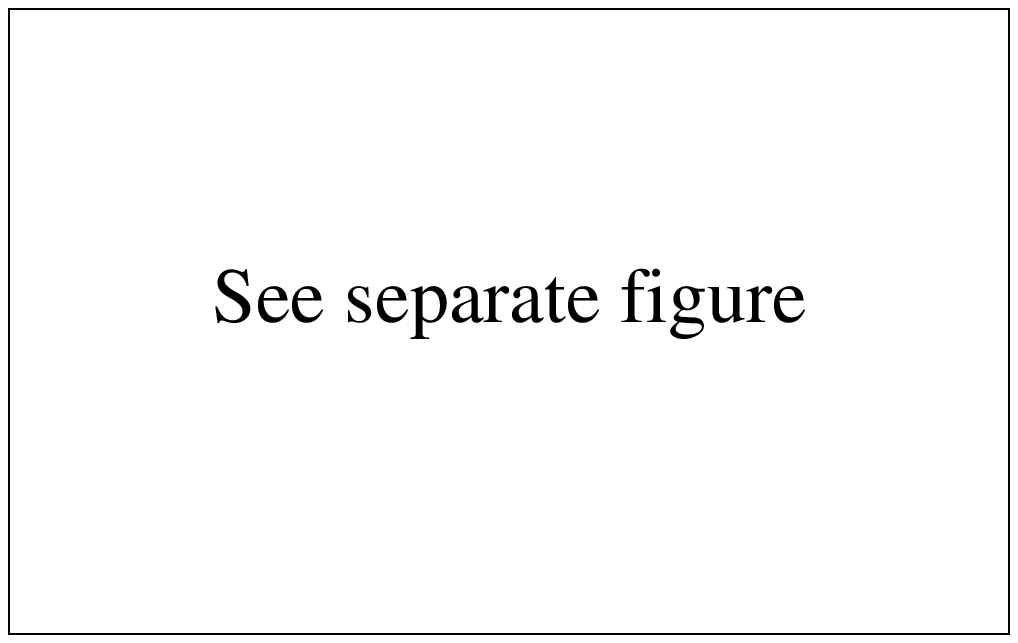,width=0.48\textwidth,angle=270}
\caption{\normalsize The upper figure maps the ratio of [OIII]$\lambda5007$/[OII]$\lambda3727$ in all fibres of IRAS F20460+1925 where [OII] is detected, for the sum of systemic and blueshifted components (see text). The lower figure maps the variation of [OIII]/H$\beta$ in the systemic component; H$\beta$ could be detected only in the individual fibres shown outlined; elsewhere, fits were carried out to the spectra of groups of fibres (not outlined). North is to the top and east to the left.} 
\label{fig:F20ionmap}
\end{figure}

The ionization structure of the gas has been derived from the [OIII]/H$\beta$ and [OIII]/[OII] ratios, since the analysis of the [NII]+H$\alpha$ complex is plagued by deblending uncertainties. Whilst [OIII]$\lambda5007$ is sufficiently strong to permit its use as a kinematic diagnostic over the region covered by the maps in Fig.~\ref{fig:F20kin}, the H$\beta$ flux in the individual fibres is poorly constrained beyond radii of $\simeq 1$ arcsec and is thus measured in groups. Fig.~\ref{fig:F20ionmap} maps the (systemic) [OIII]/H$\beta$ and [OIII]/[OII] ratios. Due to the low signal-to-noise the [OII]$\lambda 3727$ flux used here is taken from a fit with a single gaussian and, for consistency, that of [OIII] is the sum of the systemic and blueshifted components. All emission-line ratios are the observed values (ie. without correction for intrinsic extinction) unless stated otherwise. 

\begin{figure}
\psfig{figure=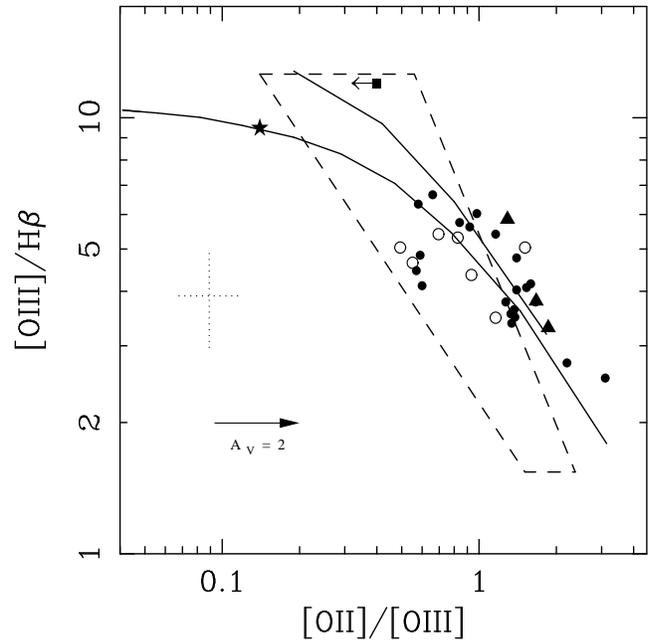,width=0.48\textwidth,angle=270}
\caption{\normalsize An ionization state diagnostic diagram for IRAS F20460+1925. The open circles represent the central seven nuclear fibres, filled circles individual fibres around the nucleus with detectable H$\beta$, the filled star the blueshifted component, the filled box the north-eastern blob and the filled triangles other groups of off-nucleus fibres, eg. the western blob. The dashed box defines the shock+precursor locus of Dopita \& Sutherland (1995) and the continuous lines delineate a series of CLOUDY photoionization models (see text for details). The plotted ratios are those observed, ie.~without correction for intrinsic reddening; the arrow denotes the correction factor for an intrinsic extinction of $A_{\rm{V}}=2$~mag (see section~4.2). The dotted cross represents a typical size for the error bars on each point, which have been omitted for clarity.}
\label{fig:F20iondiag}
\end{figure}

In Fig.~\ref{fig:F20iondiag}, the line ratios from these fibres are shown along with points representing the blueshifted component (as determined from a fit to a group of seven fibres centred on the nucleus), the kinematically distinct north-eastern and western blobs, and two other regions of the extended emission to the south-east and north-west of the nucleus, respectively. Except for the point representing the blueshifted component, the systemic [OIII] and H$\beta$ fluxes are used to calculate the line ratios, as are the [OII] fluxes for the seven nuclear fibres, in which decomposition into two velocity components was possible. It can be seen that the blueshifted component is more highly ionized than the systemic, and that the north-eastern blob is also at a high excitation level.  

The purpose of producing Fig.~\ref{fig:F20iondiag} is to use it in conjunction with the results of ionization models to understand how the line emission is powered and also to examine how the ionization state may vary across the system. The two continuous lines on this figure were obtained from a grid of CLOUDY (version 90, Ferland 1996) models using the `table agn' continuum. They represent models in which the hydrogen nucleon density at the exposed face of the (ionization-bounded) clouds assumes values of $10^{0.5}$ and $10^{3.5} \pcm$, respectively, and along which the ionization parameter $U$ varies between 10$^{-3.5}$ and 10$^{-1}$. The `table agn' continuum in CLOUDY is similar to that observed by Mathews and Ferland (1987) for a typical radio quiet active galaxy, and as such it may differ from any photoionizing continuum in this object, concerning the true shape of which we are largely ignorant. At the higher density, the models stopped due to the build-up of radiation pressure for $U>10^{-2.75}$ and are thus not shown in Fig.~\ref{fig:F20iondiag}. The dashed lines encompass the predictions of the shock + prescursor ionization models of Dopita and Sutherland (1995), with the shock velocity increasing from 200 to $500 \kmps$ vertically and magnetic parameter ($B/n^{0.5}$) in the range 0-4 G cm$^{-1.5}$. 

Whilst a single diagnostic diagram cannot be used to discriminate between the two ionization mechanisms, Fig.~\ref{fig:F20iondiag} does not even suggest the dominance of one over the other: it could be that there is a mixture of unresolved velocity components at different ionization levels. Indeed, pure photoionization by an AGN continuum seems unlikely in view of X-ray data which show that the AGN continuum source is obscured by a column density of {\em at least} $N_{\rm{H}} \simeq 2.6 \times 10^{22} \psqcm$ (Ogasaka et al. 1997); an optical-UV continuum of the requisite strength could not emerge through such obscuration. In addition, if the measured line-widths are entirely kinematic in origin, shocks would be expected to be an important source of ionization. It should, however, be noted that the shock models of Dopita \& Sutherland do not extend to velocities in excess of $500 \kmps$, whereas we measure a {\em FWHM} of around twice this value. No obvious correlation between the kinematics of the gas and its ionization state (such as would exist for shock excitation) is apparent. Neither is it possible at this signal-to-noise to measure other shock diagnostic line ratios (eg. HeII$\lambda4686$/H$\beta$ or [NeV]$\lambda3426$/H$\beta$) and thus to self-consistently implicate shock excitation. All that can be deduced with confidence from the ARGUS data is that ordinary O and B stars are not an important source of ionization. 

We comment here on the appearance of the [OIII]/[OII] map in Fig.~\ref{fig:F20ionmap}. The increase in the ratio north-east of the nucleus may result from the escape of nuclear continuum radiation along a relatively unobscured line of sight. It is interesting to speculate on the association of this structure with the north-eastern blob, which does not appear on the [OIII]/[OII] map of Fig.~\ref{fig:F20ionmap} (a $3 \sigma$ lower-limit on this quantity therein is 2.5), but which stands out clearly as a region of high excitation in [OIII]/H$\beta$ ($= 12 \pm 4$) at the same position angle. We wish to determine whether the north-eastern blob could be powered by continuum radiation from the active nucleus, or whether some in situ ionization source is required.

To address this question, we compare the rate at which the nuclear continuum supplies hydrogen-ionizing photons to the blob, to the rates required by CLOUDY models which are able to account for its observed [OIII]/H$\beta$ value and dereddened [OIII]$\lambda5007$ luminosity ($1.34 \times 10^{42}$\ergps). For the nuclear continuum we extrapolate the X-ray spectrum found by Ogasaka et al.~(1997) with a power-law of photon index 2. We do not normalize it to the value of $2.7 \times 10^{44}$\ergps~which the latter authors deduced for the absorption-corrected 2--10\keV~luminosity, because, due to scattering, the true nuclear luminosity may be some unknown factor, $f_{scatt}$, larger. Assuming the blob to lie at its projected radius of approximately 8\kpc~and to present an area $A_{cloud}$\kpcsq~of exposed cloud-face to the incident continuum, we integrate the power-law between $\lambda_{1}=228$\AA~and $\lambda_{2}=912$\AA~(ie. between the Lyman limit and the energy beyond which photons are used to ionize HeII rather than HI), to deduce that the active nucleus supplies the blob with hydrogen-ionizing photons at a rate $N_{ion}^{(1)}=7.2 \times 10^{51} A_{cloud}f_{scatt}$\ps. For given values of $f_{scatt}$ and $A_{cloud}$, the latter estimate of $N_{ion}^{(1)}$ is likely to be lower limit because the true spectrum may have a blue-bump which exceeds the extrapolation of the X-ray power law.  

The second part of the calculation involves using the CLOUDY model in conjuction with the observed blob line luminosity to calculate $A_{cloud}$, and from this the model-predicted rate, $N_{ion}^{(2)}$, at which hydrogen-ionizing photons are absorbed. For the case of ionization-bounded clouds, the $N_{ion}^{(2)}$ thus calculated is constant, regardless of the assumed values of gas density and $f_{scatt}$ in the CLOUDY model; this follows from the Zanstra method (Osterbrock 1989) which dictates that each absorbed hydrogen-ionizing photon leads to the production of one Balmer line photon, and the H$\beta$ luminosity of the blob is known observationally. Thus for ionization-bounded clouds, setting $N_{ion}^{(2)}=N_{ion}^{(1)}$ leads to the relation $A_{cloud}=36/f_{scatt}$. The projected area of the blob as we define it is $\sim 18$\kpcsq, so on the assumption that it has at most a similar extent along the line of sight and is probably composed of individual cloudlets with a covering factor of less than unity, we take $A_{cloud}<18$, which implies $f_{scatt}>2$. Given that Ogasaka et al.~(1997) deduce that the small observed value of $L_{X}/L_{Bol}$ suggests $f_{scatt} \sim 10$, the conclusion to be drawn from this calculation is that we cannot rule out that the blob is ionized by continuum radiation from the active nucleus. [Models with matter-bounded clouds do not absorb all of the incident ionizing continuum, and lead to larger values of $A_{cloud}$ for a given $f_{scatt}$, and vice versa].  

\subsection{The ISIS data}

A high resolution long-slit spectrum of this object was obtained for a number of reasons. Firstly, and as mentioned at the end of section 3.1.1., the parameters extracted from the fits to the [OIII]+H$\beta$ complex of the ARGUS data could be subject to a host of systematic effects. Secondly, the inability to decompose the [NII]+H$\alpha$ profile prohibits the determination of any intrinsic extinction. Finally, the deterioration of the signal shortward of H$\beta$ prevents the measurement of additional line ratios which could aid in the discrimination between shock+precursor ionization and AGN-photoionization. The ISIS long slit was oriented at a position angle of 30 degrees, in order to include a knot of emission which was noticed in an archival HST WFPC2 V-band image of the object (PID 5463). At a projected nuclear distance of 30\kpc~this feature could be the tidal remnant of a merger. 

\begin{figure}
\psfig{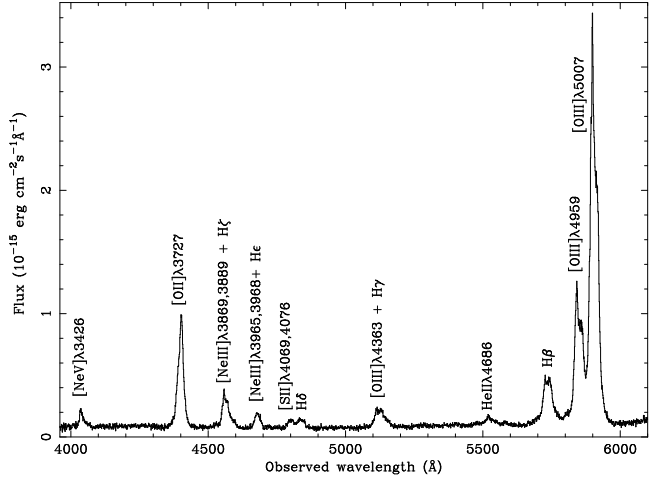}
\caption{\normalsize The nuclear spectrum of IRAS F20460+1925 taken with the ISIS spectrograph on the WHT.}
\label{fig:F20isisnuc}
\end{figure}

The nuclear ISIS spectrum is shown in Fig.~\ref{fig:F20isisnuc}; the two velocity components to the [OIII]+H$\beta$ profile are clearly resolved, with the blueshifted one being the more intense and the more highly ionized. A two component fit yields velocity widths of $870 \pm 40$ and $1010 \pm 20$ \kmps {\em FWHM} for the systemic and blueshifted components respectively, and a separation of $\simeq 990 $\kmps between them. The similarity in these widths means that the parameters obtained from the common-width model fits to the ARGUS data will not suffer seriously from any of the systematic effects described in Appendix A. This is borne out by the fact the velocity offsets of the blueshifted components derived from the ARGUS and ISIS data are in near-agreement, as are the [OIII]/H$\beta$ and [OIII]/[OII] ratios for the two components. The latter also gives independent confirmation of the ARGUS flux calibration blueward of H$\beta$. 

The spatial variation in the intensity of the blueshifted component along the direction of the slit is consistent with its being spatially unresolved, thus confirming what was suggested by the ARGUS data. Given that the new data were taken under seeing of $\simeq 1$ arcsec, it can be deduced that the true surface brightness profile of the blueshifted component is peaked within a radius of 0.5~arcsec [$\equiv2\kpc$] 

The internal extinction was estimated using two-component fits to the H$\gamma$ emission line. The velocity widths and separation of the two components were fixed at the values obtained from the [OIII]+H$\beta$ fits. When compared with the theoretical case B value for an electron temperature of $10^{4} \K$, the H$\beta$/H$\gamma$ value is consistent with an extinction of $A_{\rm{V}} \simeq 1.8$~mag for both the systemic and blueshifted components in the nuclear spectrum. Off nucleus, the systemic H$\beta$/H$\gamma$ is, within the errors, consistent with case B and therefore with no extinction, whilst that of the blue component remains close to the nuclear value (consistent with its being unresolved). For comparison, Frogel et al.~(1989) deduced an extinction of $A_{\rm{V}} \simeq 2.4$~mag from the optical narrow lines, and Veilleux et al.~(1997) used their Pa$\alpha$ flux (broad+narrow line) and the narrow line H$\alpha$ flux to compute an upper limit to the extinction to the NLR of $A_{\rm{V}}=4.6$~mag. In the diagnostic diagram of Fig.~\ref{fig:F20iondiag}, we show the correction factor which must be applied to the observed [OII]/[OIII] from the ARGUS data in order to correct for $A_{\rm{V}}=2$~mag. 

In addition to the Balmer lines and those of [OII] and [OIII], the following emission lines could be discerned in the nuclear spectrum: HeII$\lambda$4686, [SII]$\lambda$4069,4076, [NeIII]$\lambda$3965,3976 blended with H$\epsilon$, [NeIII]$\lambda$3869,3889 blended with H$\zeta$, and [NeV]$\lambda$3426. Because of blending and the presence of multiple velocity components, velocity decomposition was attempted for HeII and [NeV] only. The position of the former is consistent with its being produced predominantly by the blueshifted component and where it could be detected (within $\simeq 0.5$ arcsec of the nucleus) the HeII/H$\beta$ ratios were found to be $0.24 \pm 0.08$ and $0.14 \pm 0.05$ for the blueshifted and systemic components respectively. The blueshifted component thus lies at the same position (in terms of the values of the shock velocity and magnetic parameter) in both Fig.~\ref{fig:F20iondiag} and the HeII/H$\beta$ versus [OIII]/H$\beta$ diagnostic diagram of Dopita \& Sutherland (1995) for shock+precursor ionization. This constitutes self-consistent evidence for a shock+precursor ionization mechanism for the blueshifted component, although the implied model shock velocity of around $500$ \kmps is at variance with the observed kinematics. The [NeV]$\lambda3426$ line could be resolved into systemic and blueshifted components, having [NeV]/H$\beta$ values of $0.085 \pm 0.03$ and $0.28 \pm 0.1$, respectively. When an upward extinction correction factor of 2 is applied, the value for the blueshifted component is just compatible with the value of 0.39 predicted by the highest velocity (500\kmps) shock+precursor model of Dopita \& Sutherland (1995) for some fiducial value of the magnetic parameter (they do not present a diagnostic diagram involving [NeV]). 

The [OIII]/H$\beta$, [OII]/[OIII] and HeII/H$\beta$ values for the systemic component on nucleus satisfactorily match the previously discussed CLOUDY models for ionization parameters $U \simeq 10^{-3}$. The observed strength of [NeV]$\lambda3426$ is, however, much stronger than these models predict, but this is a well known problem with the nuclear spectra of active galaxies (Binette, Courvoisier \& Robinson~1987; Binette, Robinson \& Courvoisier~1987). We note that it can be solved by appealing to continua with additional high-energy ionizing radiation (above the Ne$^{3+}$ ionization potential of 97\eV), or to the presence of some matter-bounded clouds which produce stronger lines from the more ionized species.

Finally in connection with this object, we mention that no emission (neither line nor continuum) was seen at the position of the knot noticed in the HST image and with which the slit was aligned.

\subsection{Summary of the ARGUS and ISIS data on IRAS F20460+1925}
The ARGUS data revealed that the gas kinematics in this object are violent and complex. These findings were confirmed by the higher resolution ISIS data which showed that there are two velocity components to the [OIII]+H$\beta$ complex in the nuclear region, with {\em FWHM} of $870 \pm 40$ and $1010 \pm 20$ \kmps, separated by $\simeq 990 $\kmps. The blueshifted of the two appears to be spatially unresolved and could be due to an outflow of some kind, perhaps similar to the starburst-driven `superwinds' seen in some lower luminosity IRAS galaxies or to the minor axis outflows seen in at least a quarter of Seyfert galaxies (Colbert et al.~1996). With the exception of two areas of redshifted emission to the west and north-east of the nucleus, the systemic velocity field is otherwise relatively devoid of structure. 

The ionization state of gas at the systemic velocity is more consistent with photoionization by nuclear continuum radiation than by shocks and their precursors. The blueshifted component is more highly ionized and its emission-line intensity ratios are consistent with shock+precursor ionization, although the implied shock velocity cannot easily be reconciled with the observed kinematics.
The north-eastern blob also stands out as an area of high excitation in maps of [OIII]/[OII] and [OIII]/H$\beta$ and simple modelling suggests that it could be ionized by continuum radiation from the active nucleus; it may therefore resemble the photoionization cones seen in some lower luminosity Seyfert galaxies.  

\section{RESULTS FOR IRAS F23060+0505}
IRAS F23060+0505 is the second QSO-like object in the sample. At a redshift of 0.174, its inferred 0.5--120$\mu m$ spectral energy distribution and luminosity ($2.9 \times 10^{46}$\ergps) are characteristic of a Seyfert 1 galaxy, but its optical emission line spectrum more closely resembles that of a LINER or Seyfert 2 (Hill et al.~1987; Hough et al.~1991). Various authors have utilised near-IR/optical spectroscopy and polarimetry to model its properties as a hidden Seyfert 1 nucleus: Hough et al.~(1991) accounted for the wavelength dependence of the polarization with a dichroic model in which the nucleus and broad-line region (BLR) are viewed directly through aligned dust grains. Subsequent spectropolarimetry by Young et al.~(1996b) led to a replacement model in which light from the BLR is scattered towards us by electrons in a cone with a half-opening angle of 45 degrees, inclined at 50 degrees to the line of sight. The scattered radiation undergoes extinction equivalent to $A_{\rm{V}}=2.9$~mag, with excess flux in the near-IR accounted for by a dichroic view to the near-IR-emitting region through $A_{\rm{V}}=20$~mag. The direct view to the BLR is obscured by $A_{\rm{V}}=30 \pm 4$~mag. Hines~(1991) and Veilleux et al.~(1997) detected broad Pa$\alpha$ with ${\em FWHM} \sim 2900$\kmps, and derived lower limits on the visual extinction to the BLR of $A_{\rm{V}}>4.4$ and $>7.4$~mag, respectively. 

The findings of Young et al.~(1996b) were supported by the {\em ASCA} 2--10~keV spectroscopy of IRAS F23060+0505 reported by Brandt et al.~(1997). The latter authors demonstrated that the data were fitted best by a `two light-path model' in which the received X-rays comprise those coming directly to us through a column density of $N_{\rm{H}} \simeq 8 \times 10^{22}\psqcm$ and an additional electron-scattered component (itself absorbed by $N_{\rm{H}} \simeq 5 \times 10^{21}\psqcm$). The covering fraction of the obscuring torus (as seen from the central source) was deduced to be 99.2 per cent. The unabsorbed spectrum is a power-law with a photon index of 2 and there is evidence for the presence of an iron K-line, both of which betray the presence of an AGN. 

The host galaxy was imaged by Hutchings \& Neff~(1988) and appears to be moderately blue, indicative of a relatively young stellar population. They suggest that the presence of a tidal tail composed mainly of old red stars attests to its being in the early stages of tidal disruption. Their VLA snapshot revealed that the system contains an unresolved radio source with a 20cm flux of $6.7 \pm 0.3$~mJy. In contrast, Heisler et al.~(1994) noted that the source is unresolved at 6cm but resolved at 20cm with fluxes of 3.0 and 13.5mJy, respectively (but they do not comment on the nature of the resolved structure). 

\subsection{The ARGUS data}
\subsubsection{The line-emitting gas: its kinematics and morphology}

\begin{figure}
\psfig{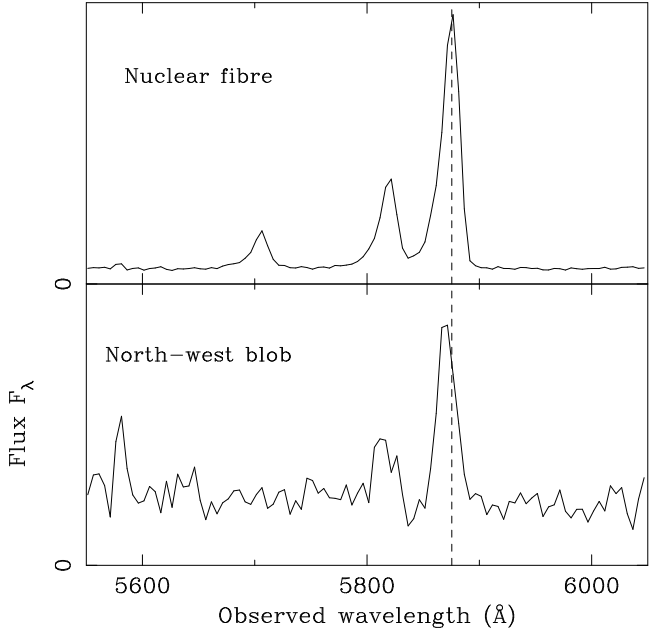}
\caption{\normalsize The [OIII]+H$\beta$ complex in the nuclear fibre and the north-western blob in IRAS F23060+1925. The dashed line indicates the position of the systemic component in the nuclear fibre.}
\label{fig:F23argus}
\end{figure}

Inspection of the CCD frames showed narrow line profiles asymmetrically {\em skewed} towards the blue, as reported by Young et al.~(1996b) [cf. IRAS F20460+1925, with its {\em two distinct, relatively symmetrical,} components]. There is thus less physical justification for the fitting of a two component model for this object than there was for IRAS F20460+1925 (still less, one in which both components have the same width). In the nucleus, the second component is blueshifted by $\simeq 830 \kmps$ with respect to the systemic and has $\simeq 20$ per cent of the latter's flux in [OIII]$\lambda 5007$. Fig.~\ref{fig:F23argus} shows the [OIII]+H$\beta$ complex in both the nucleus and a secondary enhancement in the blueshifted component (hereafter the {\em north-western blob}).

\begin{figure}
\psfig{figure=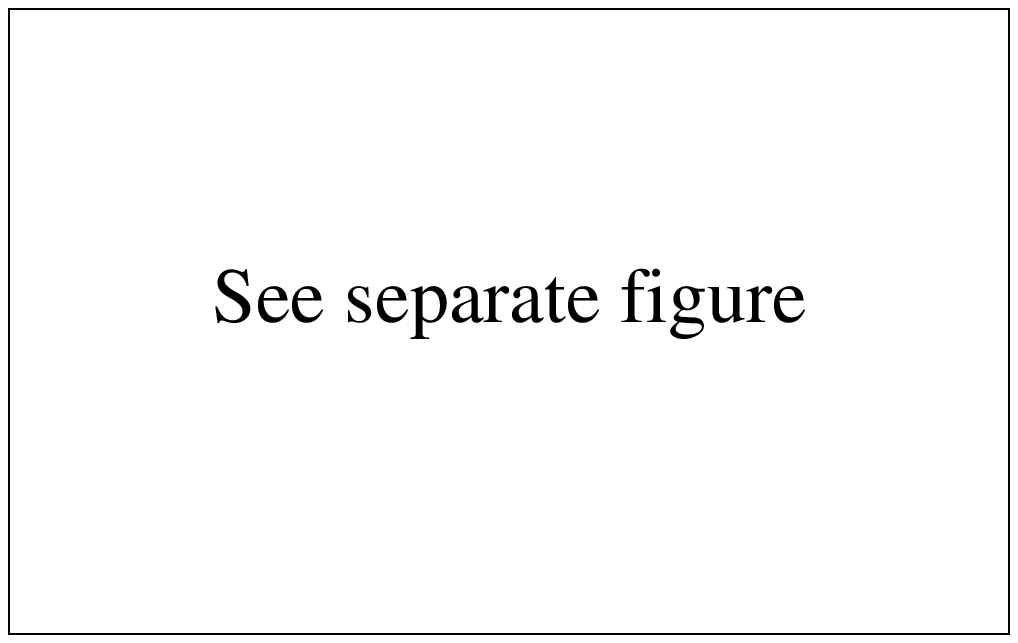,width=0.48\textwidth,angle=270}
\caption{\normalsize Reconstructed images of IRAS F23060+0505 in the following emission lines: systemic and blueshifted components of [OIII]$\lambda$5007, narrow and broad H$\alpha$, and [OII]$\lambda$3727 (see text for details). The square root of the integrated line flux has been plotted for each fibre, in order to bring out more clearly the secondary enhancement in the blueshifted component of [OIII] 7.5 \kpc~to the north-west of the nucleus (the north-west blob, as indicated). The scale is in units of $10^{-8}$ erg$^{0.5}$cm$^{-1}$s$^{-0.5}$. Discs are drawn only for those fibres with a significant line detection, and the actual flux ranges spanned by the plotted fibres are (in units of $10^{-15}$\ergpcmsqps): 0.0033--4.1 (for systemic [OIII]), 0.0026--0.87 (for blueshifted [OIII]), 0.03--0.77 (for [OII]), 0.13--2.8 (for broad H$\alpha$) and 0.013--2.2 (for narrow H$\alpha$). North is at the top and east to the left.} 
\label{fig:F23inten}
\end{figure}

Fig.~\ref{fig:F23inten} shows reconstructed images of the object in various continuum-subtracted emission lines. The surface brightness distribution of the systemic [OIII]$\lambda5007$ is centrally peaked with an approximately circular morphology, and there are two separate areas of blueshifted [OIII] emission. The first peaks on the same (nuclear) fibre as the systemic with a surface-brightness profile which follows the instrumental PSF and is thus unresolved (this {\em fitted} component may not, however, be a {\em physically distinct} velocity component). The second region of blueshifted component emission is offset from the nucleus by 7.5\kpc~to the north-west. The total [OIII]$\lambda5007$ luminosity of the system is $7.6 \times 10^{42} \ergps$, 18 per cent of which is contributed by the blueshifted component.

The [OII]$\lambda3727$ fluxes were derived from fits to this line with single-component gaussians. The decomposition of the H$\alpha$+[NII] complex could be performed and was modelled with single component narrow lines and broad H$\alpha$, the width of the latter being fixed at the value 4550\kmps~obtained by Young et al.~(1996b), owing to the contamination of its blue-wing by atmospheric absorption. The morphology of the narrow-line H$\alpha$ emission map differs from that of [OIII]$\lambda5007$, the former being significantly more extended along a north-south axis.

\begin{figure}
\psfig{figure=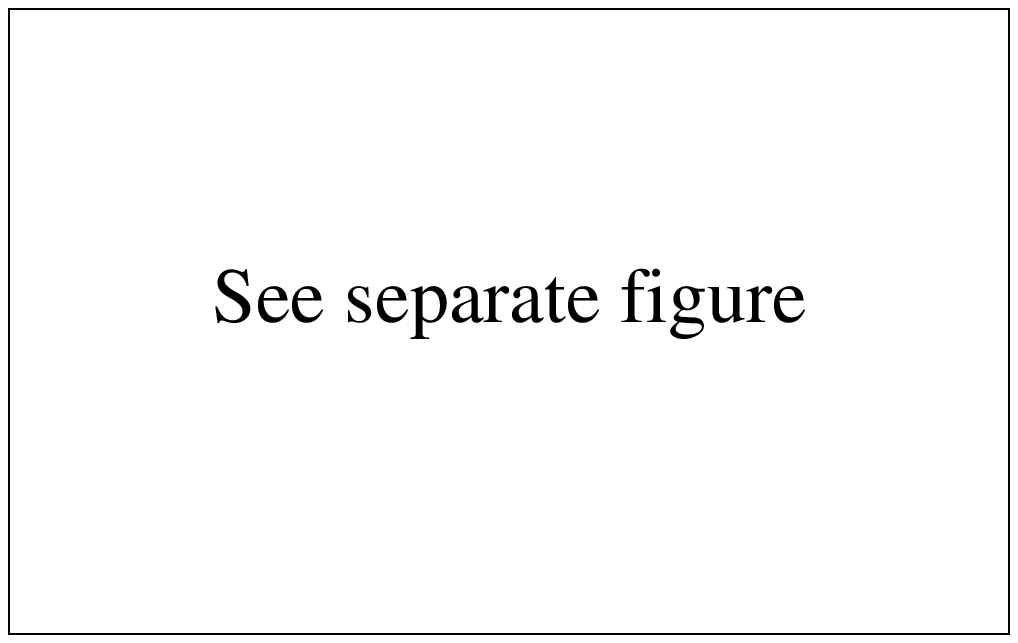,width=0.48\textwidth,angle=270}
\caption{\normalsize The upper map shows the {\em FWHM} common to the systemic and blueshifted components in a two-component fit to the [OIII]+H$\beta$ complex of IRAS F23060+0505; the linewidth was unresolved in those fibres marked with asterisks (see text). The lower map shows the radial velocity of the systemic component of [OIII]+H$\beta$ emission relative to the nuclear fibre (marked with a cross). The dashed line denotes the axis along which the velocity cut of Fig.~\ref{fig:F23vcut} is taken. North is at the top and east to the left.}
\label{fig:F23kin}
\end{figure}

\begin{figure}
\psfig{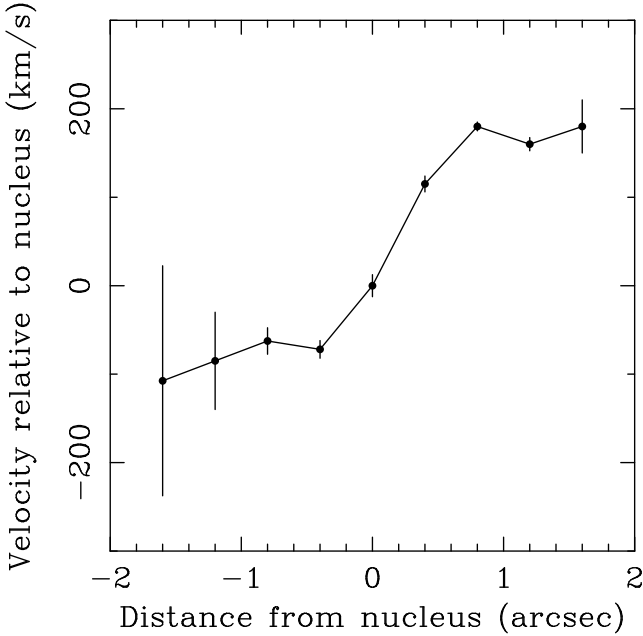}
\caption{\normalsize Radial velocity of the systemic component of [OIII] with respect to the nuclear fibre of IRAS F23060+0505, along the dashed line of Fig.~\ref{fig:F23kin}}
\label{fig:F23vcut}
\end{figure}

Fig.~\ref{fig:F23kin} shows the velocity of the systemic component of the [OIII]$\lambda5007$ emission relative to the nucleus, and its fitted {\em FWHM}. Dipolar structure is apparent in the systemic velocity field, and that it resembles normal galactic rotation is evident from Fig.~\ref{fig:F23vcut}, in which we show a velocity cut along the axis indicated by the dashed line of Fig.~\ref{fig:F23kin}. The rotation curve is asymmetric, in the sense that the kinematic centre of the galaxy is not coincident with the nucleus (defined as the peak in the [OIII]$\lambda5007$ emission). This will be returned to in section~4.2. The region of increased line-width ({\em FWHM}$\sim 600 \kmps$, as compared with $\sim 400 \kmps$ for the rest of the nebula) lies within, but is not exactly co-spatial with, the blueshifted half of the dipolar structure. This suggests that the increased line width is {\em not} a correlated response of our two-component model to the dipolar velocity structure (or vice-versa). 

In the interpretation of the fitted velocity field, we refer to the discussion of the profile-fitting systematics presented in Appendix A. The simulations described therein were performed in order to investigate how the parameters extracted from fits with a two component, common-width model depend upon the true functional form of the profile, in which the widths of the two components may be quite different from each other. The results confirm that the dipolar structure in the systemic velocity is real (indeed, it could be seen on the raw CCD frame). In contrast, the fitted relative intensity and velocity offset poorly represent the properties of any second component. That the increased {\em FWHM} might in part be a fitting artefact (ie. not entirely due to a genuine increase in the {\em FWHM} of the systemic) was suggested by its coincidence with a contigious group of fibres in which the model significantly underpredicts the flux observed between the [OIII]$\lambda5007$ and $\lambda4959$ lines. The correlation of these two features was seen during the simulations to result separately from the following changes to the true functional form of the profiles (assuming that the blue wing to the line is produced by a second gaussian component, much broader than the systemic):

(a) A fibre-to-fibre variation in the {\em FWHM} of the second component. This would imply that the latter is spatially resolved, but would {\em not} contradict the finding that the second {\em fitted} component is spatially unresolved because its fitted relative intensity significantly underpredicts the true value. It would, however, be hard to understand how a {\em spatially resolved} outflow could acquire a sufficiently large width ({\em FWHM}$\simeq1350\kmps$) for this effect to be observable. An unresolved outflow naturally generates a broad profile by integration over the spatial extent of the outflow with its spread in line of sight velocity.

\begin{figure}
\psfig{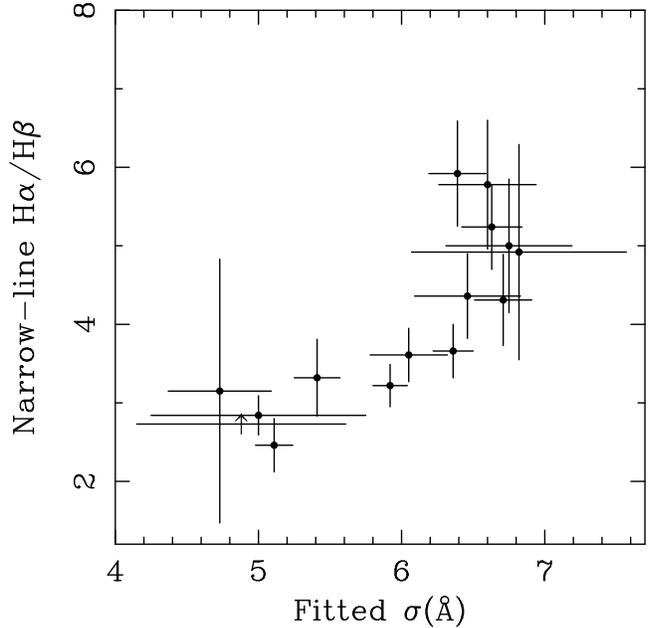}
\caption{\normalsize Narrow line Balmer decrement plotted against the fitted line width for groups of fibres in IRAS F23060+0505. The arrow denotes a three sigma lower limit for a group of fibres in which H$\beta$ is not detected}
\label{fig:F23Baldecsigma}
\end{figure}

(b) A spatial variation in the relative intensities of the two components. This could arise if the blueshifted component were due to an unresolved nuclear outflow, superimposed on the spatially-resolved systemic. Fibre-to-fibre variations in the systemic flux could be intrinsic to the line-emitting gas or produced by preferential dust-obscuration. Indeed, an archival HST WFPC2 image of the object (PID 5463) shows a prominent dust lane to the east of the nucleus, coincident with the region of interest. That this dust is in the form of a foreground screen or mixed with the line-emitting gas is confirmed by the narrow-line H$\alpha$/H$\beta$ ratio. Fig.~\ref{fig:F23Baldecsigma} shows the Balmer decrement plotted against the fitted $\sigma$~for groups of fibres spread across the nebula: a clear correlation is apparent, suggesting that this systematic effect may be operating.  

Thus, it cannot be ruled out that the observed increase in the fitted {\em FWHM} is a systematic effect, as described in (b) above, and not entirely due to a genuine increase in the velocity dispersion of either the systemic or blueshifted component. This conclusion is insensitive to whether or not the second component is spatially resolved, which in any case is difficult to ascertain from the present data. We emphasise, however, that the correlation in Fig.~\ref{fig:F23Baldecsigma} may indicate the genuine co-existence of a higher velocity dispersion with increased extinction. We shall return to this issue in the discussion of the ISIS data in section~4.2. 

We discuss here the secondary enhancement in the blueshifted component to the [OIII] at the north-western blob. At this position it is the stronger of the two velocity components, with a flux in [OIII]$\lambda 5007$ of $(4.6 \pm 1.3) \times 10^{-17} \ergpcmsqps$. It is blueshifted relative to the nuclear systemic emission by $400 \pm 50 \kmps$ and a $3 \sigma$ lower limit on [OIII]/H$\beta$ is 4.7. In view of the poor signal-to-noise and doubts about whether the blueshifted component centred on the nucleus is spatially resolved or not, it is difficult to ascertain whether this feature is a discrete `blob' or merely a region of enhanced surface-brightness in a more extensive blueshifted component. The kinematics are not helpful in this respect either, due to the uncertainty in the fitted velocity offset of the second component in fibres beyond the nucleus and the complications introduced by the fitting systematics. In the HST V-band image the position of this `blob' is seen to coincide with a local {\em deficit} in the surface-brightness. The morphology of the gas/stars around this `hole' suggests that some kind of disturbance has recently occurred there. 

\subsubsection{Ionization structure}

\begin{figure}
\psfig{figure=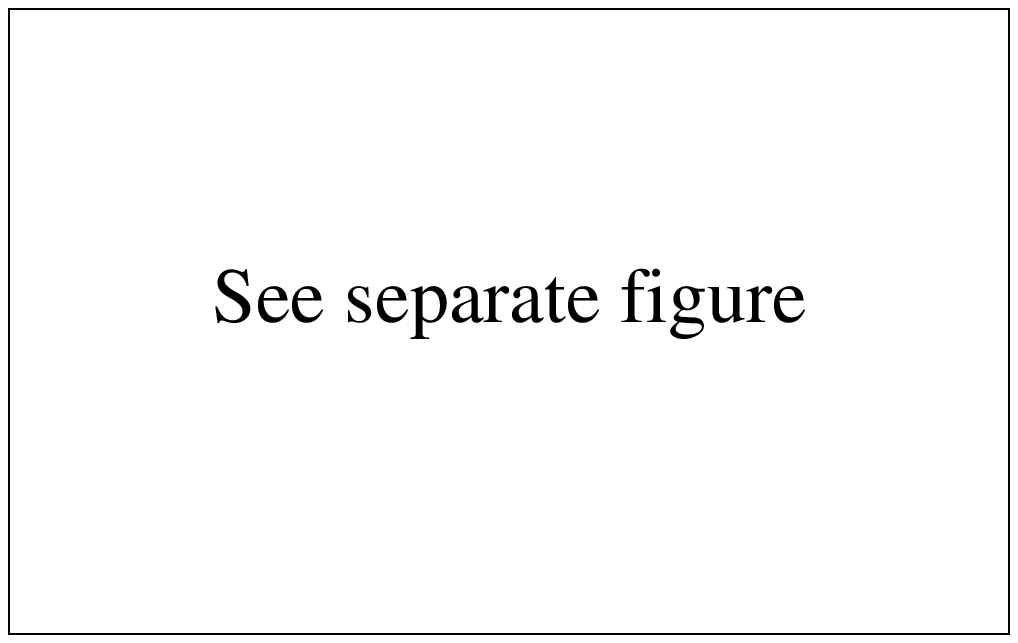,width=0.48\textwidth,angle=270}
\caption{\normalsize The upper figure shows [OIII]$\lambda5007$/[OII]$\lambda3727$ for all fibres with detectable [OII] in IRAS F23060+0505, see text for details. The lower figure shows [OIII]/H$\beta$ (systemic component), with the outlined fibres being the only individual fibres in with detectable H$\beta$; elsewhere fits were also carried out to groups of fibres (not outlined), with diamonds indicating 3$\sigma$ lower limits.}
\label{fig:F23ionmap}
\end{figure}

Maps depicting the spatial variation in the line ratios [OIII]/[OII] and [OIII]/H$\beta$ across IRAS F23060+0505 are presented in Fig.~\ref{fig:F23ionmap}. The [OII]$\lambda3727$ flux is that derived from a single gaussian fit to the line. The [OIII]$\lambda5007$ flux for this map is taken to be the systemic component alone, because if, as is likely, the blue wing is more highly ionized than the systemic, its contribution to [OII] will be much less than its $\sim 20$ per cent relative strength at [OIII].  	

The map of the systemic [OIII]/H$\beta$ suggests some anisotropy in the excitation of the gas, with the ratio being highest west and north-west of the nucleus, in the vicinity of the north-western blob. If this gas is photoionized by the nuclear continuum source, this could be due to the combined effects of an ionization cone (produced by collimation of the radiation field within the nuclear regions) and a general patchiness in the distribution of obscuring material in the galaxy itself (as the fact that the apex of the `cone' is not centred on the nucleus would imply). The latter may also account for the offset between the kinematic centre of the galaxy and the observed position of the nucleus. 

\begin{figure}
\psfig{figure=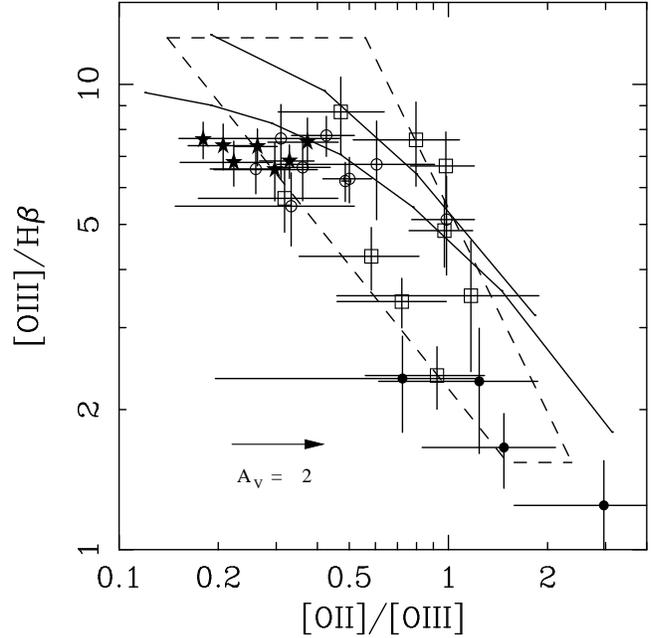,width=0.48\textwidth,angle=270}
\caption{\normalsize [OIII]/H$\beta$ versus [OIII]/[OII] for the systemic emission components of the fibres in IRAS F23060+0505 where all the relevant lines are detected. The photoionization and shock+precursor loci and the extinction-correction arrow are as in Fig.~\ref{fig:F20iondiag}. Star symbols denote the seven central nuclear fibres, open circles fibres at a radius of 0.8 arcsec, open squares fibres at 1.2 arcsec radius and filled circles some fibres at radii $\geq 1.5$ arcsec.}
\label{fig:F23iondiag}
\end{figure}

In Fig.~\ref{fig:F23iondiag} we show a line ratio diagnostic diagram for the individual fibres in Fig.~\ref{fig:F23ionmap}. The role of nuclear photoionization is again difficult to assess because of our ignorance of the shape of the optical-UV continuum. Of the two model loci shown in Fig.~\ref{fig:F23iondiag}, that corresponding to shock+precursor ionization provides the closer description of the ionization state. In the context of this model, fibres at greater radii are consistent with lower shock speeds. It is difficult to test shock+precursor ionization by looking for the expected power-law dependence of the H$\beta$ flux on the shock-velocity because systematic effects in the spectral fitting and projection effects undoubtedly mean that the fitted {\em FWHM} in a particular fibre does not directly indicate the speed of the shock responsible for the line emission. Nevertheless, shock excitation remains attractive, not least because X-ray and spectropolarimetric studies indicate that the nuclear continuum source is almost completely covered with thick obscuring material. As in section~3.1, we mention, however, that a single diagnostic diagram cannot be used to rigorously discriminate between the two ionization mechanisms. 

\subsection{The ISIS data}
In acquiring an ISIS spectrum of this object we sought a better understanding of the origin of the increased {\em FWHM} and of the nature of the blue-asymmetry in the line profiles, as well as a greater number of emission-line diagnostics. A slit position angle of 135 degrees was chosen, in order to cover both halves of the dipolar velocity field and the north-western blob.

\begin{figure}
\psfig{figure=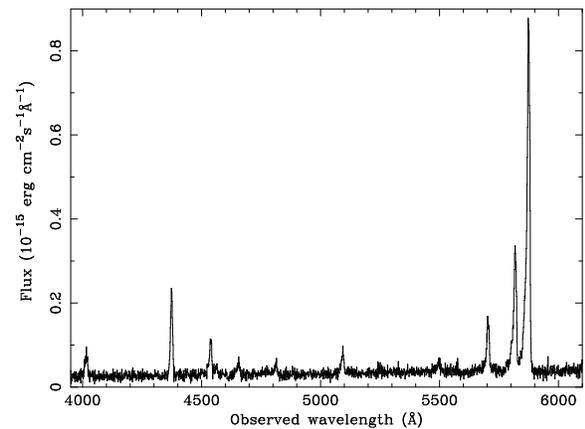,width=0.48\textwidth,angle=270}
\caption{\normalsize The nuclear ISIS spectrum of IRAS F23060+0505.}
\label{fig:F23isisnuc}
\end{figure}

\begin{figure}
\psfig{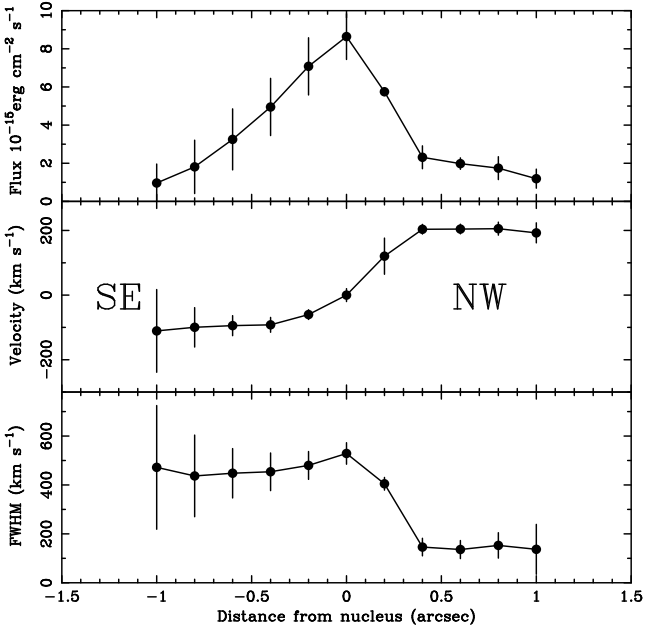}
\caption{\normalsize Variation of the fitted model parameters with position along the ISIS slit for the systemic component in IRAS F23060+1925. From top to bottom are shown the [OIII]$\lambda5007$ flux, the velocity with respect to the nucleus and the {\em FWHM}. The north-western and south-eastern ends of the slit are indicated.}
\label{fig:F23isispar}
\end{figure}

\begin{figure}
\psfig{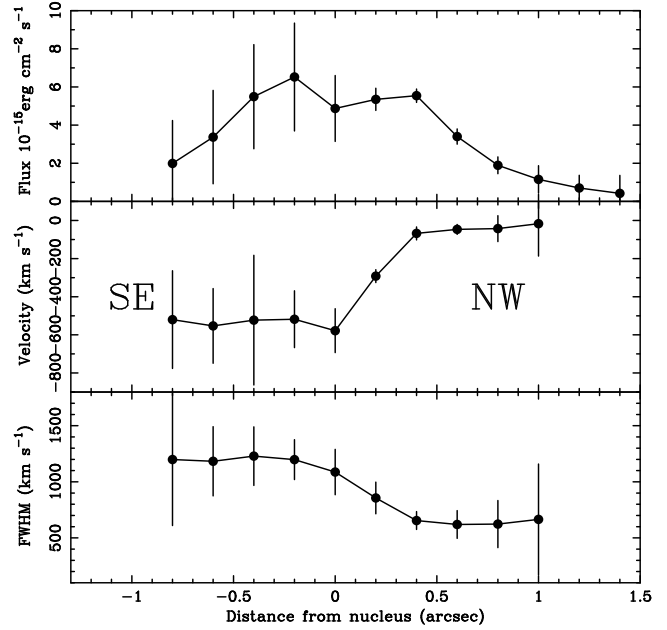}
\caption{\normalsize As in figure \ref{fig:F23isispar} but for the blueshifted component. The velocity is relative to the nuclear systemic.}
\label{fig:F23isisparbs}
\end{figure}

\begin{figure}
\psfig{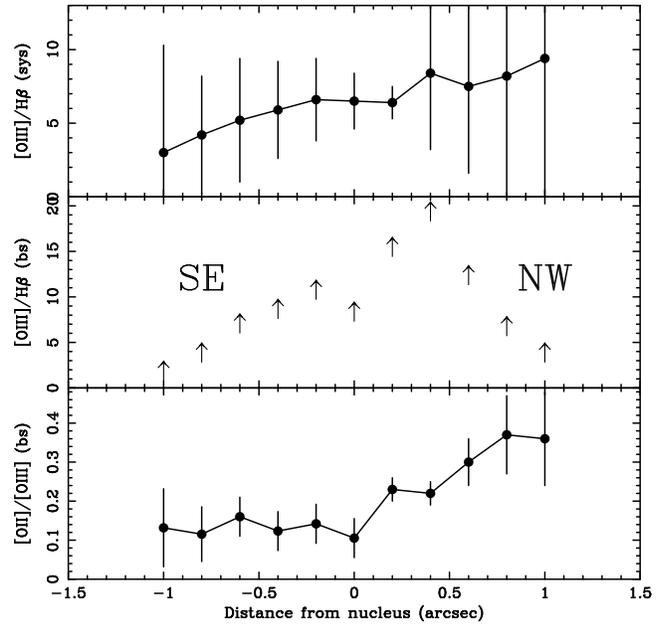}
\caption{\normalsize From top to bottom are shown the [OIII]/H$\beta$ ratios of the systemic and blueshifted components and [OII]/[OIII] for the blueshifted component, with arrows denoting three sigma lower limits for spectra with a questionable second component to H$\beta$.}
\label{fig:F23isisionbs}
\end{figure}

Fig.\ref{fig:F23isisnuc} shows the nuclear ISIS spectrum, with its blue-asymmetric line profiles. Spectra along the slit were fitted with the two component model of the [OIII]+H$\beta$ complex (with the two widths no longer constrained to be equal) and the variations in the fitted parameters are shown in Figs.~\ref{fig:F23isispar}, \ref{fig:F23isisparbs} and \ref{fig:F23isisionbs}. The principal findings are that:

(1) The asymmetry in the systemic velocity rotation curve seen in Fig.~\ref{fig:F23vcut} is confirmed.

(2) The width of the systemic is constant on the SE half of the dipolar structure, and declines to a much smaller value on the NW half.

(3) The velocity offset of the blueshifted component from the nuclear systemic is constant to the SE of the nucleus and decreases in amplitude towards to the NW. 

(4) The width of the blueshifted component decreases from SE to NW. 

(5) The [OII]/[OIII] ratio is constant to the SE of the nucleus, but increases NW of it.   

We thus conclude that, if the blue-asymmetric line profiles are due to a single second component -- and for the purpose of ascertaining whether this is the case the ISIS data are no more useful than the ARGUS data -- it is poorly modelled by the common-width two-component model used to fit the latter. The revision of some of the conclusions of section 4.1 is thus necessary. Firstly, it can no longer be maintained that any second component is spatially unresolved. Secondly, the systematic effects which were cited as possible causes for the region of increased {\em FWHM} in Fig.~\ref{fig:F23kin} are probably not primarily responsible for it, the ISIS data showing instead that there are genuine increases in the widths of both the systemic and blueshifted components. 

We discuss now the ionization state of the gas, as revealed by the ISIS data. The measured [OIII]/H$\beta$ and [OII]/[OIII] values for the systemic component accord well with the ARGUS values, despite the systematic effects in the fits to the latter. Some information on the ionization state of the blueshifted component, which could not be reliably measured using the ARGUS data, is presented in Fig.~\ref{fig:F23isisionbs}; as the presence of a second component to H$\beta$ is questionable, we show 3 sigma lower limits in the [OIII]/H$\beta$ plot. The [OII]/[OIII] ratio is approximately constant in the blueshifted half of the dipolar velocity field, but increases with nuclear distance on the other half of this structure. 

Whilst many of the other lines present in the IRAS F20460+1925 spectrum also appear in that of IRAS F23060+0505, the analysis of them is here less straightforward. This is because the signal-to-noise ratio is lower, and the disparity in their widths means that the blueshifted wing falls below the noise level before the systemic part of the line. The smaller separation of the two components also makes it difficult to attribute unambiguously a line which can be fitted with only a single component to one or the other of the two. A two component fit to HeII$\lambda4686$ shows that the blueshifted component is marginally stronger, and yields HeII/H$\beta$ $\simeq 0.1-0.23$ and (lower limits of) $\simeq 0.3-0.9$, respectively, for the systemic and blueshifted components. Similarly, a fit to [NeV]$\lambda3426$ is dominated by the blueshifted component and yields [NeV]/H$\beta$ $\simeq 0.1-0.18$ and (lower limits of) $\simeq 0.5-1.1$, respectively, for the two components. The systemic HeII/H$\beta$ and [NeV]/H$\beta$ values are, when the latter is extinction-corrected, consistent with the shock+precursor models of Dopita \& Sutherland (1995) (as the ARGUS data suggested), with velocities up to $\sim 500$\kmps. The latter range matches that shown in Fig.~\ref{fig:F23isispar} for the systemic {\em FWHM}.

The observed line-ratios of the blueshifted component are less-easily understood; the measured [OII]/[OIII] and [OIII]/H$\beta$ ratios do not discriminate between the two ionization mechanisms and the HeII/[OIII] and [NeV]/[OII] ratios fail to match the predictions of the otherwise plausible CLOUDY models or the values given by Dopita \& Sutherland (1995) for the highest velocity shock+precursor models. Given, however, that the latter is for 500\kmps~whereas the {\em FWHM} of the blueshifted component is much larger (Fig.~\ref{fig:F23isisparbs}), it seems more likely that the gas is ionized by high velocity shocks (with velocities $\sim 1000$ \kmps) than by the non-thermal continuum of a central active nucleus. 

The H$\beta$/H$\gamma$ ratio for the blueshifted component is poorly constrained, and that of the systemic is consistent with an extinction of $A_{\rm{V}} \simeq 1.2$ mag on nucleus. Fits to average spectra in the red- and blueshifted halves of the dipolar structure yield extinctions of $A_{\rm{V}} \simeq 0.4$ and 1.1, respectively, although the extinction on the blueshifted half is not uniform, the Balmer decrement being in places consistent with the case B value. Nevertheless, that the extinction is generally higher on the blueshifted half of the structure is consistent with the findings of the ARGUS data and with the location of the dust lane in the HST image of this object. For comparison, Hough et al.~(1991) deduced $A_{\rm{V}}=3.4 \pm 0.5$ mag from single component fits to the narrow H$\alpha$ and H$\beta$ lines, and Veilleux et al.~(1997) used their combined broad+narrow Pa$\alpha$ flux to estimate an upper limit on the NLR extinction of $A_{\rm{V}}<5.5$~mag.

Although the slit was aligned to pass through the secondary enhancement in the blueshifted component noticed in the ARGUS data, the noise level in the ISIS data is such that nothing more can be added to the discussion of section~4.1.1.

\subsection{A kinematic model for IRAS F23060+0505}
In the context of the merger-induced formation scenario, we speculate here on an interpretation of this object's kinematics. The long tidal tail (Hutchings \& Neff~1988) signals an on-going merger in which the galaxies are near to the final coalescence. If not the result of dust obscuration, the asymmetric rotation curve may be due to the merger-induced disturbance of the pre-encounter velocity field of the principal progenitor galaxy; the increased {\em FWHM} and dust lane in the same region may have a similar origin. The kinematics of the blueshifted component shown in Fig.~\ref{fig:F23isisparbs} are suggestive of some sort of bipolar nuclear outflow. Within the ULIRG evolutionary paradigm, the latter would be ascribed to a starburst-driven superwind, but we also note that the spatial structure of the [OIII] profiles may be related to the interaction of the previously-noted radio structure with the ambient medium, similar to that found by Whittle et al.~(1988) for a number of Seyfert galaxies (see section~7.2). 

\subsection{Summary of the ARGUS and ISIS data on IRAS F23060+0505}
The systemic velocity field of this object has a dipolar structure with an axis along a position angle of 30 degrees. The rotation curve is, however, asymmetric, in the sense that it has amplitudes of 100 and 200\kmps~on its blueshifted and redshifted halves, respectively. The systemic {\em FWHM} is approximately constant at 475\kmps~in the SE half, but declines towards the NW. The second velocity component with which the blue asymmetry in the lines of the [OIII]+H$\beta$ complex was modelled is much broader than the systemic ({\em FWHM} up to 1200 \kmps) and offset from the latter by -600\kmps~on nucleus. A secondary enhancement to the blueshifted velocity component is seen 7.5\kpc~NW of the nucleus, but the data do not allow us to ascertain whether it is a discrete `blob' of emission or a region of enhanced surface-brightness in a more extensive blueshifted component. It would be of interest to obtain a high-resolution radio map of this object, to examine the relationship between the spatial structure of the [OIII] profiles and any extended radio emission.  

The [OIII]/H$\beta$ and [OIII]/[OII] values from the ARGUS data suggest that shocks and their precursors ionize the systemic velocity component. Other diagnostics from the ISIS data confirm this, and show that the implied shock velocities match the observed systemic {\em FWHM}. The ionization state of the blueshifted component is not consistent with the CLOUDY model used throughout this paper, nor with the highest velocity (500\kmps) shock+precursor model of Dopita \& Sutherland (1995). Since the blueshifted {\em FWHM} is much greater than 500\kmps~and studies at other wavelengths indicate that the active nucleus is very heavily obscured, we also favour shocks (perhaps faster than those which Dopita \& Sutherland compute) as the ionization mechanism for the blueshifted component. 

\section{JKT multi-band images of IRAS F20460+1925 and F23060+0505}
The ARGUS and ISIS spectroscopic data have enabled us to study in some detail the kinematics and ionization structure of the ionized gas in these systems. Further clues to their evolutionary history are provided by the colours of the host galaxy stellar populations. Hutchings and Neff (1988) compared the B- and R-band images of IRAS F23060+0505 and observed a series of arms or curved luminous regions extending in all directions from the nucleus, which is itself resolved at radii of $<1$ arcsec. They observed also a series of knots or secondary nuclei, particularly in B light, at distances of 2 to 4 arcsec from the nucleus, and claimed that these colour differences distinguish regions of recent massive star-formation from older stars. For IRAS F20460+1925, Frogel et al. (1989) obtained images in the V- and I-bands but did not report any colour gradients across the galaxy, which has an absorption-corrected $M_{\rm{V}}=-22.3$, comparable to that for giant spiral or elliptical galaxies. They did, however, draw attention to the knot of V-band emission 6.5 arcsec to the south-west of the nucleus (seen also in the HST V-band image, as mentioned in section 3.1) and deduced that it, or any other companion object, must be at least 3 magnitudes fainter in I than the IRAS source. 

Using the JKT data, we confirm the complex colour structure of IRAS F23060+0505 reported by Hutchings \& Neff, the most prominent features being the knot of B-band emission in the plume to the SW and a ``fan-like'' R-band extension of the nucleus in roughly the same direction. The host galaxy of IRAS F20460+1925 is most apparent in the I-band and exhibits no discernable colour gradients which, along with its quasi-elliptical surface-brightness profile, point to a relatively ancient stellar population. The V-band emission knot is absent from all three of the present filter images and from the ISIS spectrum, suggesting that its SED is peaked strongly in the V-band. Whilst these facts may be used as a basis for speculation about its properties, it should be borne in mind that, lacking a knowledge of its redshift, its physical association with the main galaxy remains no more than a hypothesis.   

\section{THE LOWER LUMINOSITY OBJECTS}
With bolometric luminosities close to $10^{13}$ \Lsun, IRAS F20460+1925 and F23060+0505 straddle the conventionally-adopted ultra/hyperluminous divide. In this section, we discuss the ARGUS observations of two less powerful objects [$L_{\rm{FIR}}=10^{12.2}$~and~$10^{12.4}$ \Lsun], included in the programme for comparison with the QSO-like systems. As mentioned in section~2.1, the seeing progressively deteriorated during this period, thus precluding a comparison on the terms which had been originally envisaged. Whilst the study of the {\em spatial structure} of the kinematics and ionization state is slightly impaired, integrated properties such as the total line luminosities can still be extracted. 

\subsection{IRAS F01217+0122}
IRAS F01217+0122 lies at z=0.137 and was included in the long-slit spectroscopic sample of Armus, Heckman \& Miley (1989) (hereafter AHM), who commented upon its broad, blue-asymmetric emission line profiles. An R-band image shows two apparently interacting galaxies, but spectroscopy reveals that the larger of these is a foreground object (Armus, Heckman \& Miley 1990). 

Because of the poor seeing, the ARGUS spectra were analysed in hexagonal groups of seven fibres out to a radius of 1.6 arcsec from the nucleus, beyond which no emission line flux could be detected. The [OIII]+H$\beta$ complex in these spectra were fitted with the two-component common-width model, with a series of four defective CCD columns on the blue-wing of the [OIII]$\lambda4959$ line being masked out (but this does not adversely affect the fitted parameter values). The kinematics appear to be almost entirely devoid of structure: the systemic velocity and fitted {\em FWHM} are substantially constant across the nebula, the latter assuming a value of $\simeq 1050$ \kmps. The velocity offset of the spatially-unresolved blueshifted component is $\simeq 1320$ \kmps. It may be the case that the observed kinematics are dominated by the effects of the seeing on a centrally-peaked surface-brightness profile, and that they are thus a poor representation of the true velocity field. The total luminosity of the system in [OIII]$\lambda 5007$ is $3.53 \times 10^{42}$~\ergps, 13 per cent of which comes from the blueshifted component.   

\begin{figure}
\psfig{figure=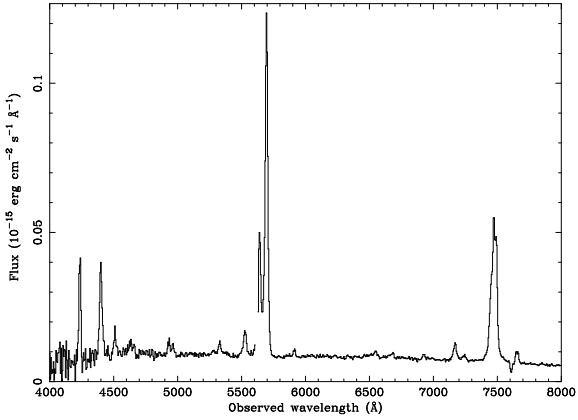,width=0.48\textwidth,angle=270}
\caption{\normalsize A portion of the nuclear spectrum of IRAS F01217+0122. Note the strength of [NeIII]$\lambda 3869$. A series of bad columns on the blue-wing of the [OIII]$\lambda4959$ line was masked out, as described in the text. }
\label{fig:F012nuc}
\end{figure}

The nuclear spectrum (shown in Fig.~\ref{fig:F012nuc}) exhibits a number of strong emission lines blueward of H$\beta$, including [OII]$\lambda 3727$, [NeIII]$\lambda 3869$, [OIII]$\lambda 4363$ and HeII$\lambda 4686$ which, when fitted with single component gaussian profiles, yield line ratios of [OII]/[OIII]$\lambda5007 =0.25 \pm 0.03$, [NeIII]/[OII]$=1.0 \pm 0.16$, HeII/H$\beta = 0.53 \pm 0.12 $ and [OIII]$\lambda 4363$/[OIII]$\lambda 5007=0.019 \pm 0.01$ (the [OIII]$\lambda 5007$ and H$\beta$ fluxes being those of the systemic component). Note the unusually strong [NeIII]$\lambda 3869$ and [OIII]$\lambda 4363$; the latter equates to an electron temperature of $\sim 22000$\K~after correction for an extinction of $A_{\rm{V}} \simeq 3.6$~mag (deduced from the Balmer decrement $F_{\rm{H\alpha}}/F_{\rm{H\beta}}=9.46$ given by AHM). The fitted value of [OIII]/H$\beta = 14.2 \pm 2.3$ on nucleus is consistent with the value measured by AHM, which was the largest of all the objects in their sample. Because there is evidence for underlying Balmer absorption in some of the off-nuclear spectra of this object (in eg. Fig~\ref{fig:F012specC}), the nuclear spectrum may also suffer some contamination. When comparing the observed emission line intensity ratios with those predicted by models, we thus restrict our attention to those not involving H$\beta$. The large value of [NeIII]$\lambda3869$/[OII]$\lambda3727$ strongly favours AGN photoionization over any of the shock+precursor of Dopita \& Sutherland (1995); in particular, the CLOUDY model shown in Fig.~\ref{fig:F20iondiag} with a hydrogen nucleon density of $10^{0.5}$\pcmcu~and ionization parameter $U\simeq 10^{-1.5}$, is able to reproduce the observed value of this ratio, whilst also being consistent with the measured [OIII]$\lambda4363$/HeII$\lambda4686$ and [OIII]$\lambda4363$/[OIII]$\lambda5007$ values. The ionization state of the blueshifted component cannot be constrained at the signal-to-noise level of these data. 

\begin{figure}
\psfig{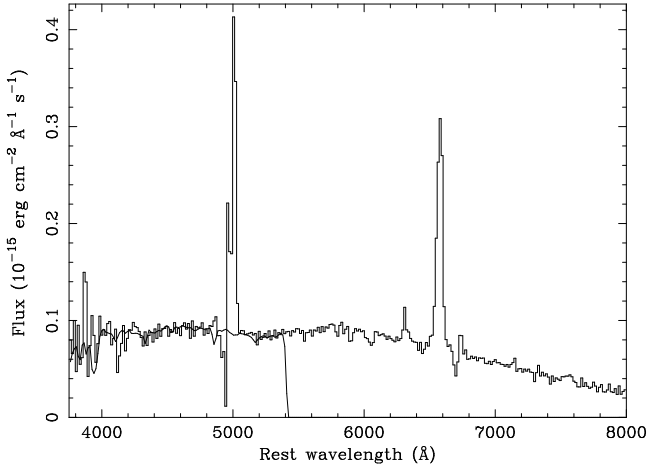}
\caption{\normalsize The spectrum of a group of fibres 1.2 arcsec south of the nucleus of IRAS F01217+0122, plotted at a dispersion of 15\AA~per pixel on a rest-wavelength scale. The best-fit stellar template model is plotted up to 5400\AA, and dominated by F5 stars. The strong emission lines were assigned zero statistical weight in the fit.}
\label{fig:F012specA}
\end{figure}

\begin{figure}
\psfig{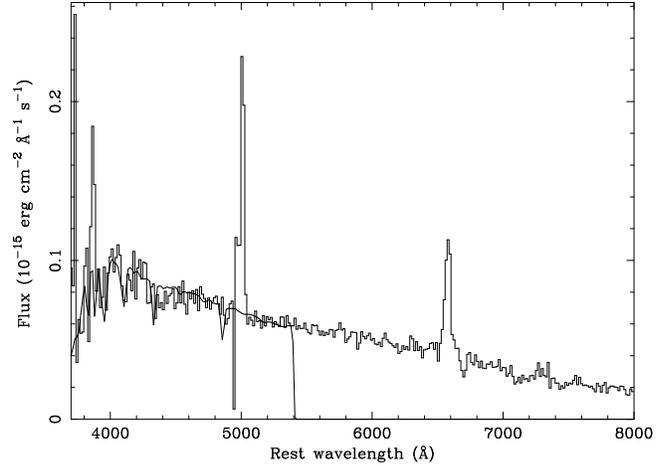}
\caption{\normalsize As in Fig.~\ref{fig:F012specA} but for a group of fibres 1.2 arcsec north-west of the nucleus. The stellar model is dominated by A5 stars with a lesser contribution from those of class F5.}
\label{fig:F012specC}
\end{figure}

That there may have been a relatively recent ($<1$ Gyr), localised episode of star-formation is also suggested by the existence of a colour gradient across the galaxy. To quantify the strength of the blue continuum in excess of that from a giant elliptical galaxy (which, we note, may not be appropriate for the underlying host galaxy) we fitted the spectra with a series of templates representing O5,~B5,~A5,~F5 and G5 stellar models (Kurucz 1979). The spectra of the fibre groups were placed on a rest wavelength scale and changed to a dispersion of 15\AA~per pixel to improve the signal-to-noise ratio, but were not corrected for intrinsic reddening, lacking as we do a reliable measure of the latter (because of the Balmer line absorption). Fitting was performed between 3600--5400\AA~(beyond which the template spectra are not tabulated) with strong emission lines being assigned zero statistical weight. As the smoothed galaxy spectra exhibit prominent H$\gamma$ and H$\delta$ absorption lines, the fitted models are not surprisingly dominated by stars of spectral types A and F. Figs.~\ref{fig:F012specA} and~\ref{fig:F012specC} compare the fitted spectra of two groups of off-nucleus fibres. The excess blue light of the entire galaxy (compared with an old elliptical galaxy stellar population) can essentially be reproduced with $\sim (1.8 \pm 0.8) \times 10^{5}$ A5 stars and $\sim (1.4 \pm 1.1) \times 10^{10}$ F5 stars. The large errors, however, prevent this result from being used to infer significant information about the star-formation history of the galaxy. Given that the main sequence lifetimes of O5,~B5,~A5 and F5 stars are $\sim 3.7 \times 10^{6}$, $3.9 \times 10^{7}$, $6.0 \times 10^{8}$ and $3.3 \times 10^{9}$ yr, respectively (quoted by Shapiro \& Teukolsky 1983), it suffices to say that the results are consistent with an intermediate age ($\sim 1$Gyr) stellar population. 

\subsection{IRAS F01003-2238}
IRAS F01003-2238 lies at a redshift of 0.118 and has attracted the attention of numerous authors since its inclusion in the original sample of warm ULIRGs by Sanders et al.~(1988b). Its popularity stems from its classification as a Wolf-Rayet galaxy by Armus, Heckman \& Miley (1988) (hereafter AHMII). They interpreted a band of features at a rest wavelength of 4660\AA~and other broad but weaker features as arising from $\simeq 10^{5}$ late Wolf-Rayet stars of the W-N sub-type. More recently, Surace et al. (1998) presented WFPC2 B- and I-band images of this and nine other warm ULIRGs. The only notable circumnuclear feature they saw was a chain of knots extending from the SE to the NW within 1 arcsec of a putative nucleus. The latter are especially prominent in the B-band and, if interpreted as regions of recent star-formation, have a mean age of about 7 Myr, consistent with the young ages ascribed to the stellar populations by AHMII. Despite being the brightest galaxy in a small group, it shows no discernable distorted structure. Near-infrared spectroscopy by Veilleux et al.~(1997) found the galaxy to have a strong component of narrow Pa$\alpha$, with a blue-asymmetric line-profile base, but they saw no evidence for a hidden BLR. 

\begin{figure}
\psfig{figure=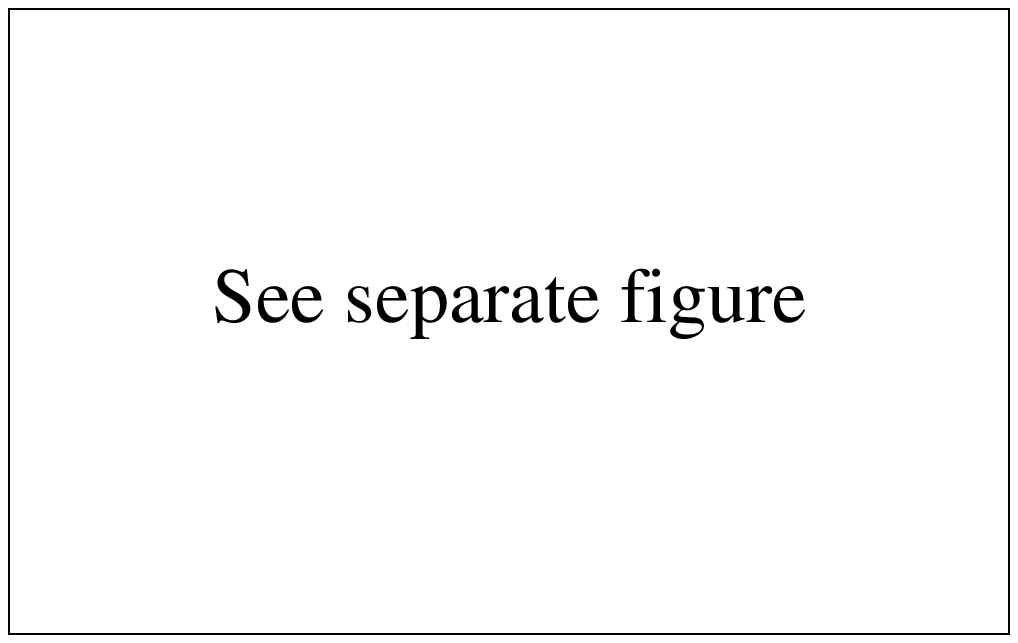,width=0.48\textwidth,angle=270}
\caption{\normalsize
\label{fig:F01003kin} 
The upper map shows the {\em FWHM} common to the systemic and blueshifted components in a two component fit to the [OIII]+H$\beta$ complex of IRAS F01003-2238, corrected for the position-dependent instrumental resolution. The fitting was carried out on hexagonal groups of seven fibres because of the poor seeing; the crosses denote the nuclear group and the asterisks mark a group in which the linewidth was unresolved. The lower map shows the radial velocity of the systemic component of [OIII]+H$\beta$ emission. North is at the top and east to the left.}
\end{figure}

The gas kinematics were deduced from fits to the [OIII]+H$\beta$ complex and are more globally structured than those of IRAS F01217+0122, showing a dipolar systemic velocity field with an amplitude of $\simeq 140$~\kmps across the galaxy. The {\em FWHM} of the lines are also broader in the blueshifted parts of this structure (as for IRAS F23060+0505), with a {\em FWHM} varying from $\simeq 450$ to $1000$~\kmps over the same region. Both these features are shown in the kinematic maps of Fig~\ref{fig:F01003kin}. The second component to the [OIII]+H$\beta$ complex used to model the blue wing to the lines is offset from the systemic by $\simeq -1150$ \kmps on nucleus, with a relative intensity in [OIII] of $\simeq 0.5$. The luminosity of the entire system in [OIII]$\lambda5007$ is $1.32 \times 10^{42}$~\ergps, of which 29 per cent is contributed by the second component.

The nuclear [OIII]/H$\beta$ ratio of this object is the lowest in the present sample, with a value of $4.1 \pm 0.8$ The only other emission lines which can be discerned blueward of H$\beta$ are [OII]$\lambda$3727, and, with slightly less than 3$\sigma$ significance, [NeIII]$\lambda 3869$. When fitted with single gaussians these yield line ratios of [NeIII]$\lambda 3869$/[OII]$\lambda3727=0.26 \pm 0.15$ and [OIII]$\lambda5007$/[OII]$\lambda3727=1.65 \pm 0.6$ (using the flux of only the systemic [OIII] component). The fitting of a two component model to the [NII]+H$\alpha$ complex yields a nuclear [NII]/H$\alpha$ ratio of $0.19 \pm 0.04$, placing the object in the HII region part of the [NII]/H$\alpha$ versus [OIII]$\lambda5007$/H$\beta$ diagnostic diagram of Veilleux \& Osterbrock~(1987). The low signal-to-noise of the present data precludes the identification of the Wolf-Rayet features noted by AHMII and the determination of the ionization state of the blueshifted velocity component.  

\begin{figure}
\psfig{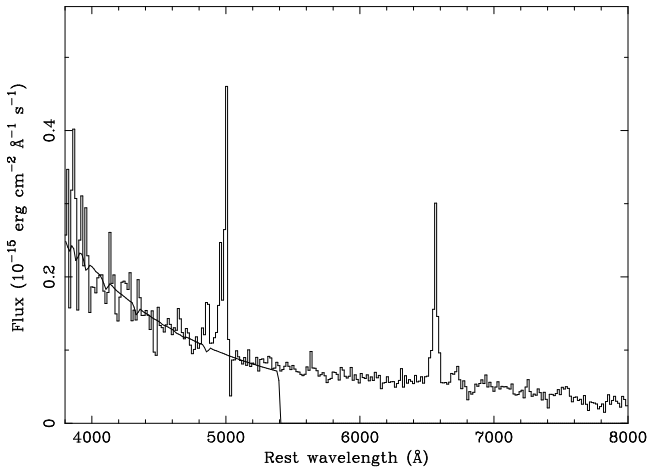}
\caption{\normalsize The spectrum of the group of fibres in IRAS F01003-2238 approximately coincident with the star-clusters identified by Surace et al.~(1998), plotted at a dispersion of 15\AA~per pixel on a rest-wavelength scale. The best-fit stellar template spectrum, which is dominated by 05 stars, is plotted up to 5400\AA. The strong emission lines were assigned zero statistical weight in the fit.}  
\label{fig:F010clusters}
\end{figure}

Due to the poor seeing ({\em FWHM}$\simeq2$~arcsec), the four knot-like star clusters and the putative nucleus identified by Surace et al.~(1998) all fall within the seeing disk. It is thus not possible to use the ARGUS data to address the questions raised by the latter authors, concerning whether the observed Seyfert activity is associated with the putative nucleus and whether the blue wing to the emission lines originates in a superwind from the star clusters. There is, however, a marked colour gradient across the galaxy, with the spectra of groups to the north and west of the nucleus (ie. in the vicinity of the star-clusters) possessing very steeply rising blue continua, clearly indicative of a young stellar population. Stellar template spectra were used to quantify the amount of blue light using the procedure described above for IRAS F01217+0122, the only difference being that an extinction correction was applied to the IRAS F01003-2238 spectra using the measured H$\alpha$/H$\beta$ ratio. On nucleus, the observed ratio translates into $A_{V}=1.4$~mag, whilst on the star clusters it is consistent with the case B value and therefore with no extinction. To the south and east of the nucleus extinctions as high as $A_{V}=2.5$~mag are deduced. This spatial gradient in the extinction conflicts with the findings of Surace et al.~(1998), who claimed that the observed colour structure was indicative of a {\em uniform} dust screen. For comparison, Veilleux et al.~(1997) quote extinctions to the NLR derived from optical and near-infrared lines of $A_{V}=2.7$ and $5.5$~mag, respectively. 

The spectra of the star clusters are dominated by O5 and B5 stars (in terms of the fraction of the 4500\AA~continuum flux they contribute), although all but the O5 contribution is poorly constrained in the fits. If fitted solely with O5 stars, a total of $3.2 \pm 0.2 \times 10^{5}$ such stars are required to account for these spectra. Fig.~\ref{fig:F010clusters} shows the fit of the stellar template model to the spectrum of the two groups of fibres to the north and west of the nucleus, in the vicinity of the star-clusters. Considering that the main sequence lifetime of an O5 star is $\sim 3.6 \times 10^{6}$~yr, this result is consistent with the age of 10~Myr deduced by Surace et al.~(1998) for the oldest of these clusters by analysing evolutionary tracks on an M$_{\rm{B'}}$ versus B-I diagram. When extinction corrections are applied to the spectra of the other groups of fibres, they too are seen to have similarly strong blue continua. This suggests that star formation has occurred recently throughout the galaxy, and not just in the visible clusters where the extinction is observed to be consistent with $A_{V}=0.$ On the assumption that the putative nucleus identified by Surace et al.~(1998) is in fact a compact starburst, we find that the extinction-corrected blue continuum of the entire galaxy can be accounted for with $9.2 \pm 0.5 \times 10^{6}$ O5 stars. Taking $10^{5.7}$\Lsun~for the bolometric luminosity of an O5 star (Allen 1973), this quantity of stars could plausibly account for the entire bolometric luminosity of the galaxy. This, along with the relatively low-ionization emission-line spectrum described above, may be taken to argue against the presence of an active nucleus in this object. Surace et al.~(1998) found that the position of the putative nucleus on the M$_{\rm{B'}}$ versus B-I diagram was anomalous with respect to the other objects in their sample, and inconsistent with its being a reddened QSO nucleus. 

\begin{table*}
\begin{center}
\caption{Summary of the properties of the four ULIRGS}
\begin{tabular}{llllll} \hline
  & F01003-2238 & F01217+0122 & F20460+1925 & F23060+0505\\ \hline

z: & 0.118 & 0.137 & 0.181 & 0.174 \\

log(L$_{\rm{Bol,FIR}}^{\star}/\Lsun$): & 12.2 & 12.4 & 13.2 & 12.9\\ \\

Total L[OIII]$\lambda$5007   &      &      &     &        \\
($10^{43}$\ergps):  & 0.13 & 0.35 & 2.6 & 0.76 \\

BS$^{\dagger}$ L[OIII]$\lambda$5007 &       &       &      &         \\
($10^{43}$\ergps):         & 0.038 & 0.044 & 1.4 & 0.14 \\

Fraction of total &  &  &  &  & \\
L[OIII]$\lambda$5007 in BS:  & 0.29  & 0.13  & 0.54  & 0.18     \\ \\

Environment: & group & field & field & field \\

Merger features: & none & none & none & tidal tail to SW\\ \\

Spatial extent$^{\ddagger}$      &          &       &      &                 \\
of [OIII]$\lambda$5007 (\kpc):&   10      &    10       &   20      &   15             \\ \\

BS $\Delta$v(nuclear) (\kmps): & $-1150\pm60$ & $-1320\pm80$ & $-975\pm25$ & $-830\pm40$\\

BS morphology: & unresolved & unresolved & unresolved & extended to NW \\

BS ionization &              &                  &                   &                                \\
source: &   unconstrained    &   unconstrained  &  shock+precursor  & shock+precursor  \\       
        &                    &                  &                   &                       \\ \\

{\em FWHM} (\kmps): & 400--1000 & 1050 & 1000 & 400--600 \\ \\  

Systemic velocity field: & dipolar structure:  & structureless  & structureless   & dipolar structure: \\ 
               & $\Delta v \simeq 140$~\kmps, &                & but for two  & $\Delta v \simeq 300$~\kmps,      \\
               & increased {\em FWHM} &                & isolated blobs  & increased {\em FWHM} \\
               & on blue side &                &                 & on blueshifted side \\ \\

Systemic ionization &  HII region      &  photoionization  & photoionization  & shock+precursor  \\
source:              &  photoionization   &  by AGN           & by AGN        &                 \\ \\

Stellar        & Wolf-Rayet spectral  & A5 \& F5 stars; & no evidence    & blue knots   \\
populations:   & features (AHMII);    & 1 Gyr old burst  & for recent     & (Hutchings \& Neff~1988)       \\
	       & O5 \& B5 stars; blue &                 & star formation &      \\
               & knots $\sim 7$ Myr old &                 &                &                   \\
	       & (Surace et al.~1998)      &                 &                &                    \\ \hline
\end{tabular}
\end{center}
$\star$ See Table~2 \\
$\dagger$ BS refers to the velocity component blueshifted with respect to the systemic \\
$\ddagger$ Linear size of [OIII]$\lambda$5007 nebula in the direction of greatest extent \\
\end{table*}

\section{DISCUSSION}
Having dealt separately with each of the four ULIRGs in turn, we discuss here the comparative properties of the sample as a whole. The findings of the preceeding sections are summarised in Table 3. We aim to understand whether the observed diversity of kinematic and ionization structures {\em within our sample} (not necessarily for the class of ULIRGs as a whole) can be accommodated within some general framework, such as the evolutionary scenario of Sanders et al.~(1988a). 

\subsection{The systemic velocity fields}
Within our small sample of four objects, there appear to be two essentially different kinds of systemic velocity field. The first group comprises IRAS F20460+1925 and F01217+0122, in which the systemic velocity is approximately constant across the galaxy with a uniformly high {\em FWHM} ($\sim 1000$ \kmps); in contrast, objects in the second group (IRAS F23060+0505 and F01003-2238) exhibit more organised velocity fields, with dipolar structure and a {\em FWHM} which is smaller than in objects in the previous group but which is largest on the blueshifted half of the dipolar structure. In the context of the merger scenario for ULIRGs, we speculate that objects in the first group are old mergers in which the pre-encounter velocity structure has been destroyed by merger-induced turbulence, whilst those in the second are much younger and still retain some record of the velocity structure (eg. dominantly rotational) of the progenitors. It could instead be that the kinematic disturbance is due a galactic-scale outflow (perhaps initiated by a merger) rather than to the effects of a merger {\em per se}. In a sample of this size, significance should probably not be attached to the fact that in both of the objects in the second group, the region of increased {\em FWHM} lies on the {\em blueshifted} half of the dipolar structure. 

In support of the above interpretation, we refer to the archival HST WFPC2 V-band images of IRAS F20460+1925 and F23060+0505. That of the former shows approximately elliptical isophotes and, with the
 exception of a central point source, an otherwise featureless surface 
brightness profile; a discrete knot of emission at a projected radius of 30\kpc~may be a fossil tidal remnant. In contrast, the image of IRAS F23060+0505 shows a long tidal tail extending to the south-west, a prominent dust lane and numerous bright knots of emission, all of which suggest that the system is a relatively young merger which has recently undergone prodigious star formation, as claimed by Hutchings \& Neff (1988). The JKT images (section~5) confirm these differences in the general morphology and colour structure of the two systems. Thus, IRAS F23060+0505 appears to be dynamically the younger of the two mergers (if indeed one did occur in IRAS F20460+1925), nominally consistent with the above expectations.

For the lower-luminosity objects IRAS F01217+0122 and F01003-2238 the above kinematic interpretation receives less direct support, as neither object shows obvious signs of interaction (Armus, Heckman \& Miley~1990 and Sanders et al.~1988b, respectively). This is perhaps surprising, given that a very large proportion of ULIRGs reside in interacting or merging systems (Clements et al.~1996). Nevertheless, the large {\em FWHM} do suggest that mergers or galactic-scale kinematic disturbances of some other kind (eg.~superwinds) are occurring. As both the latter are intimately connected with bursts of star formation, we can compare the times elapsed since their onset by comparing the properties of their stellar populations. In section~6.1, the rising blue continuum of IRAS F01217+0122 was modelled with a $\sim1$~Gyr old starburst, whereas that of F01003-2238 (section~6.2) was dominated by O5 and B5 stars, consistent with the age of 10~Myr deduced by Surace et al.~(1998) for the oldest star clusters. That IRAS F01003-2238 should have the younger stellar population thus conforms with the above interpretation of the kinematics. 

\subsection{The blueshifted velocity components}
The emission lines of the [OIII]+H$\beta$ complex of all the objects are blue-asymmetric in appearance. We assumed that this is due to a {\em physically distinct} velocity component blueshifted with respect to the systemic part of the line (cf. the situation in which a {\em single} mechanism produces the entire line profile). In all but one of the objects (IRAS F23060+0505) it appears to be spatially unresolved, and (where reliable measurement is possible) more highly ionized than the systemic velocity component. Emission line intensity ratios point to ionization by shocks and their precursors rather than to photoionization by an AGN continuum. Table 3 shows that the [OIII]$\lambda5007$ luminosity of the blueshifted component increases with L$_{\rm{Bol,FIR}}$ (approximately as L$_{\rm{Bol,FIR}}^{1.7}$) and that the outflow velocity on nucleus is generally higher in the less powerful systems. We caution, however, that these conclusions are potentially sensitive to the line-profile fitting systematics (Appendix A).

Considering the paucity of constraints which we have upon their properties, we can do little more than speculate about the origin of these blueshifted components. One possibility is that they are outflows driven by the collective effects of many supernovae, the so-called starburst-driven superwinds; the latter are a common feature of lower-luminosity IRAS galaxies (see Armus, Heckman \& Miley 1989, 1990) and are probably also present in other ULIRGs (eg. Veilleux et al.~1995; Kim et al.~1998). However, given the strong evidence for active nuclei in three of these objects, we should also consider {\em their} role. IRAS F20460+1925 is a case in point: unlike the other galaxies, the CCD frames of this object show a second, {\em separate} velocity component rather than a blue {\em asymmetry} to the line profile, which contributes a much greater fraction of the total [OIII]$\lambda5007$ than the blue wings of the other galaxies. And in addition, this system is likely to be an old merger with no sign of recent star formation to power a putative superwind.

As mentioned earlier, Colbert et al.~(1996) found that more than a quarter of Seyfert galaxies show evidence for large-scale minor axis galactic outflow and discussed the mechanisms which may drive them. Whittle et al.~(1988) demonstrated that [OIII] profile substructure in Seyfert galaxies is frequently associated with extended radio structure; such emission components typically have high excitation ([OIII]/H$\beta \sim 10-15$) and are undergoing systematic outflow from the galaxy.  

\subsection{Ionization mechanisms}
Emission line intensity ratios suggest that in most of the objects the gas at the systemic velocity is ionized by a hard radiation field (IRAS F01003-2238 is a possible exception -- see below). Discriminating further between photoionization by an AGN and shock+precursor ionization is generally problematic, as we are ignorant of the shape of the AGN continuum spectra and of the properties of the gas clouds. In IRAS F23060+0505 there is good evidence for shock+precursor ionization, from both the line ratios themselves and their concordance with the systemic {\em FWHM} derived from the ISIS data. As mentioned in section~4.1.2, this is consistent with indications from X-ray spectroscopy that the central engine is obscured over almost 4$\pi$ steradian by a column density of $N_{\rm{H}} \sim 10^{23} \psqcm$. The line of sight obscuration to the X-ray-emitting regions of IRAS F20460+1925 is also high, but less is known about its solid angle coverage; it is thus possibile that radiation could escape to power the line emission, which has observed line ratios consistent with AGN photoionization. Furthermore, line ratio maps of this object show that the [OIII]/[OII] ratio increases north-east of the nucleus, terminating in a region of high [OIII]/H$\beta$. On account of its low ${\em FWHM}$, we speculated that this `NE blob' could comprise gas in the host galaxy which is ionized by continuum radiation from the active nucleus, an interpretation which simple energetic considerations do not render implausible. These differences between IRAS F20460+1925 and F23060+0505 may be taken to reinforce their previously-deduced relative positions along the merger-induced `evolutionary sequence': the `photoionization cone' of the former resembles smaller-scale structures in EELRs of lower luminosity Seyfert galaxies; with its higher and more uniform (in terms of solid angle coverage) circumnuclear obscuration giving rise to shock-powered line emission, IRAS F23060+0505 is perhaps at a somewhat earlier evolutionary stage; AGN photoionization may become more important as the central regions are swept clear of obscuring material.       

Turning now to the lower luminosity objects, the high [OIII]/H$\beta$ and [NeIII]$\lambda3869$/[OII]$\lambda3727$ ratios suggest that AGN photoionization is the dominant mechanism in IRAS F01217+0122. The case of IRAS F01003-2238 is more ambiguous: it has a much lower [OIII]/H$\beta$ ratio than the other objects and on an [OIII]/H$\beta$ versus [NII]/H$\alpha$ diagnostic diagram, it falls just within the HII region area. The young stellar population could thus be just as important a source of ionization as a putative AGN. The lack of high ionization lines ([NeIII]$\lambda3869$ is detected at just less than the 3$\sigma$ level) suggests, moreover, that the former mechanism may dominate. Indeed, the number of number of O5 stars needed to account for the extinction-corrected blue continuum could plausibly power the entire bolometric luminosity of the galaxy. Thus, a comparison of the roles played by AGN and HII region photoionization in IRAS F01217+0122 and F01003-2238 is also consistent with their locations on the putative evolutionary sequence.  

\section{Conclusions}
We have presented the results of an area spectroscopic study of the galactic-scale nebulae around four ULIRGs, focusing on the kinematics and ionization structure of the line-emitting gas. We accommodated our findings within the evolutionary scenario of Sanders et al.~(1988a), by showing how the kinematics may be used to assess the merger status in a way which is consistent with other indications of the latter.

The systemic velocity fields in the four objects are of essentially two kinds: IRAS F23060+0505 and F01003-2238 exhibit orderly dipolar rotation, whilst IRAS F20460+1925 and F01217+0122 show little coherent structure in the systemic together with $\rm{{\em FWHM}}>1000$\kmps, which we claim are indicative of more advanced mergers. For IRAS F23060+0505 and F20460+1925 we support this interpretation with morphological and colour information derived from JKT and archival HST data. At lower luminosity, fits to the continuum spectra of IRAS F01217+0122 and F01003-2233 are consistent with the less advanced merger having the younger starburst population. Furthermore, the emission-line spectrum of IRAS F01217+0122 clearly reveals an AGN, whilst IRAS F01003-2238 has an HII region spectrum and could be powered entirely by the starburst: we are plausibly seeing two systems at different stages along the merger-induced transition from starburst to AGN.

\section*{ACKNOWLEDGMENTS}
CSC thanks the Royal Society for support, and RJW and RGA acknowledge PPARC for a studentship and an Advanced Fellowship, respectively. We thank Alastair Edge for taking the JKT data and Pascal Teyssandier for the development of the ARGUS-dedicated IRAF packages. The CFHT is operated by CNRS of France, NRC of Canada, and the University of Hawaii. The WHT and JKT are operated on the island of La Palma by the Isaac Newton Group in the Spanish Observatorio del Roque de los Muchachos of the Instituto de Astrofisica de Canarias.

{}

\appendix
\section{Emission-line fitting systematics}
Throughout this paper, the [OIII]+H$\beta$ profiles from the ARGUS data have been fitted with a model comprising two {\em equal width} gaussian velocity components, the spectral resolution and signal-to-noise ratio being insufficient to justify the fitting of anything more complex. Whilst such models provide a good parameterisation of the blue asymmetric line profiles, the physical interpretation of the velocity offsets and relative intensities so derived may be subject to a number of systematic effects. This is especially likely if, as found by AHM for a sample of lower-luminosity IRAS galaxies, the true profile comprises a systemic component and a {\em much broader}, blueshifted base.   

\begin{figure}
\psfig{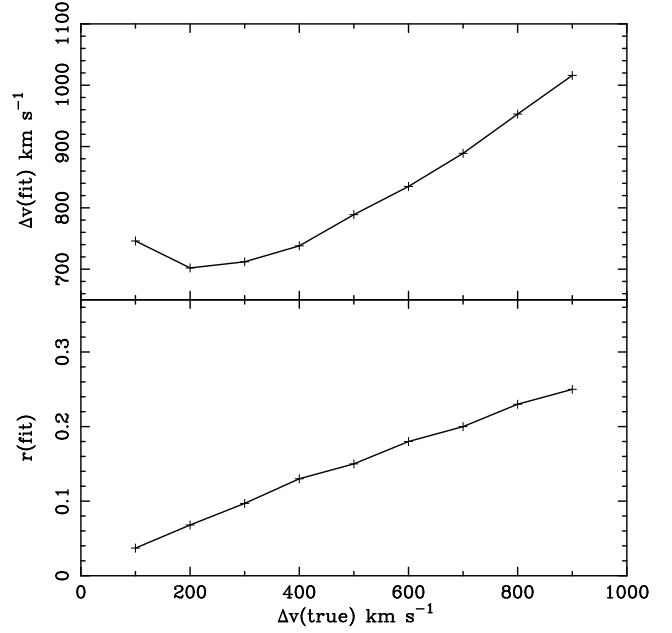}
\caption{\normalsize Plots showing the variation of the fitted relative intensity $r_{fit}$ and velocity offset $\Delta v_{fit}$ as functions of $\Delta v_{true}$, for $\sigma_{narrow}=5.5$\AA, $\sigma_{broad}=8.25$\AA~and $r_{true}=0.4$}
\label{fig:system1}
\end{figure}

To investigate these effects, we performed some simulations. Synthetic, noise-free, [OIII]+H$\beta$ profiles were generated, comprising a systemic gaussian velocity component (centroid $\lambda_{narrow}$, width $\sigma_{narrow}$) and a second, broader, component with width $\sigma_{broad}$, separated from the systemic by $\Delta v_{true}$, and with a fraction $r_{true}$ of the latter's flux (the [OIII]/H$\beta$~ratios of the two components were set equal). The profiles were then fitted with the aforementioned two-component, common-width model and the responses of the fitted parameters ($\lambda_{fit},\sigma_{fit},\Delta v_{fit}$ and $r_{fit}$) to variations in the true profile parameters were investigated. The results are as follows:\\ \\
(1) With $\sigma_{narrow}=5.5$\AA~and $\sigma_{broad}=8.25$\AA~(corresponding to {\em FWHM} of 500 and 1000\kmps~respectively, after correction for an instrumental resolution of 4.2\AA), it was found that $\sigma_{fit}$~varies little as a function of $\Delta v_{true}$~(remaining near $5.8$\AA) and that $\lambda_{fit}$~follows very closely the movement of $\lambda_{narrow}$. In contrast, the value of $r_{fit}$ declines from 0.25 to 0.03 as $\Delta v_{true}$~varies over 900 to 100\kmps~and the behaviour of $\Delta v_{fit}$~is even more unexpected: over the same range in $\Delta v_{true}$, $\Delta v_{fit}$ changes from $\simeq 1000$ to just $\simeq 700 \kmps$, as illustrated in Fig.~\ref{fig:system1}. At the highest values of $\Delta v_{true}$, the model underpredicts the flux between the [OIII]$\lambda5007$ and $\lambda4959$ lines.\\ \\

\begin{figure}
\psfig{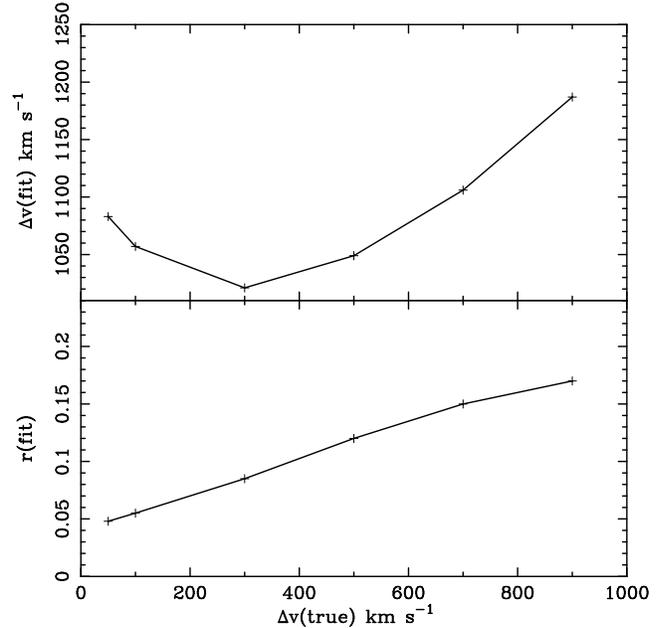}
\caption{\normalsize Plots showing the variation of the fitted relative intensity $r_{fit}$~and velocity offset $\Delta v_{fit}$~as functions of $\Delta v_{true}$, for $\sigma_{narrow}=5.5$\AA, $\sigma_{broad}=12.0$\AA~and $r_{true}=0.4$}
\label{fig:system2}
\end{figure}

(2) $\lambda_{fit},\Delta v_{fit}$ and $r_{fit}$ are insensitive to changes in $\sigma_{narrow}$.\\ \\
(3) As $\sigma_{broad}$~is increased to $12$\AA, it is observed that $\lambda_{fit}$ remains practically unchanged, while $r_{fit}$ and $\sigma_{fit}$ decrease and increase by small amounts, respectively. Once again $\Delta v_{fit}$ exhibits a strong response: as $\Delta v_{true}$ varies over 900 to 100\kmps, $\Delta v_{fit}$ ranges over $\simeq 1190$--$1050\kmps$. The flux deficit between the [OIII]~lines becomes significantly more pronounced. Fig.~\ref{fig:system2} shows $r_{fit}$ and $\Delta v_{fit}$ as functions of $\Delta v_{true}$.\\ \\

(4) As $r_{true}$ is increased from 0.4 to 0.8, $\lambda_{fit}$ and $\Delta v_{fit}$ scarcely respond, and $r_{fit}$ increases by roughly a factor of 2. The response of $\sigma_{fit}$, particularly at large values of $\sigma_{broad}$, is of interest: eg. with the latter set at $12$\AA, $\sigma_{fit}$ varies from about 6 to 6.5 as $r_{true}$ varies over the above range and the flux deficit between the [OIII]~lines remains significant.\\ \\

\begin{figure}
\psfig{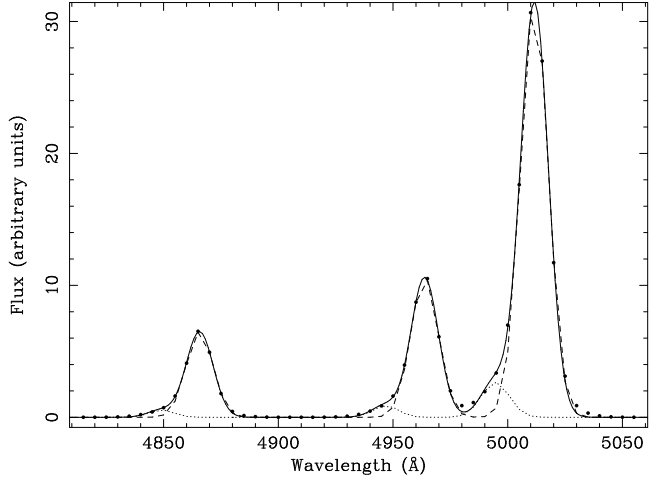}
\caption{\normalsize The dots sample the synthetic [OIII]+H$\beta$ profile with $\sigma_{broad}=12.0$\AA~and $\sigma_{narrow}=5.5$\AA, and $\Delta v_{true}=300 \kmps$. The solid line shows the best-fit two component, common-width model, with systemic (dashed) and blueshifted (dotted) components having $\sigma_{fit}=6.0$\AA, separated by $\Delta v_{fit}=1022$\kmps. Note the great disparity between the model and fitted velocity differences, and between $r_{true}=0.4$ and $r_{fit}=0.085$. A flux deficit between the [OIII]$\lambda5007$ and $\lambda4959$ lines is also visible.}      
\label{fig:system3}
\end{figure}

Fig.~\ref{fig:system3} illustrates how, even when a visually good fit to the complex is obtained with the common-width model, the extracted parameter values can poorly represent those of the true profile.

\begin{thebibliography}{}


\bibitem []{} Allen C.W., Astrophysical Quantities, Athlone Press, 1973


\bibitem []{} Armus L., Heckman T.M., Miley G.K. (AHM II), 1988, ApJ, 326, L45

\bibitem []{} Armus L., Heckman T.M., Miley G.K. (AHM), 1989, ApJ, 347, 727

\bibitem []{} Armus L., Heckman T.M., Miley G.K, 1990, ApJ, 364, 471

\bibitem []{} Binette L., Courvoisier T.J.-L., Robinson A., 1987, A\&A, 190, 29

\bibitem []{} Binette L., Robinson A., Courvoisier T.J.-L., 1987, A\&A, 194, 65 
\bibitem []{} Bohlin R.C., Savage B.D., Drake J.F., 1978, ApJ, 224, 132

\bibitem []{} Brandt W.N., Fabian A.C., Takahashi K., Fujimoto R., Yamashita A., Inoue H., Ogasaka Y., 1997, MNRAS, 290, 617

\bibitem []{} Clements D.L., Sutherland W.J., McMahon R.G., Saunders W., 1996, MNRAS, 279, 477 

\bibitem []{} Colbert E.J.M., Baum S.A., Gallimore J.F., O'Dea C.P., Lehnert M.D., Tsvetanov Z.I., Mulchaey J.S., Caganoff S., 1996, ApJS, 105, 75

\bibitem []{} Crawford C.S., Vanderriest C., 1996, MNRAS, 283, 1003

\bibitem []{} Crawford C.S., Vanderriest C., 1997, MNRAS, 285, 580

\bibitem []{} Dopita M.A., Sutherland R.S., 1995, ApJ, 455, 468

\bibitem []{} Ferland G.J., 1996, {\em Hazy, a Brief Introduction to Cloudy}, University of Kentucky Department of Physics and Astronomy Internal Report.

\bibitem []{} Frogel J.A., Gillett F.C., Terndrup D.M., Vader J.P., 1989, ApJ, 343, 672


\bibitem []{} Genzel R, et al., 1998, ApJ, 498, 579

\bibitem []{} Heckman T.M., Armus L., Miley G.K., 1990, ApJS, 74, 833


\bibitem []{} Heisler C.A., Vader J.P., 1994, AJ, 107, 35


\bibitem []{} Hill G.J., Wynn-Williams C.G., Becklin E.E., 1987, ApJ, 316, 11


\bibitem []{} Hines D.C., 1991, ApJ, 374, L9

\bibitem []{} Hines D.C., Schmidt G.D., Smith P.S., Cutri R.M., Low F.J., 1995, ApJ,450, L1


\bibitem []{} Hines D.C., Wills B.J., 1993, ApJ, 415, 82

\bibitem []{} Hough J.H., Brindle C., Wills B.J., Wills D., Bailey J., 1991, ApJ, 372, 478

\bibitem []{} Hutchings J.B., Neff S.G., 1988, AJ, 96, 1575


\bibitem []{} Kay L.E., Miller J.S., 1989, BAAS, 21, 1099

\bibitem []{} Kim D.-C., Veilleux S., Sanders D.B., 1998, ApJ, 508, 627

\bibitem []{} Kurucz R.L., 1979, ApJS, 40, 1

\bibitem []{} Lutz D., Spoon H.W.W., Rigopoulou D., Moorwood A.F.M., Genzel R., 1998, ApJ, 505, L103

\bibitem []{} Mathews W.G., Ferland G.J., 1987, ApJ, 323, 456   


\bibitem []{} Mihos J.C., Bothun G.D., 1998, ApJ, 500, 619

\bibitem []{} Ogasaka Y., Inoue H., Brandt W.N., Fabian A.C., Kii T., Nakagawa T., Fujimoto R., Otani C., 1997, PASJ, 49, 179

\bibitem []{} Osterbrock D.E., 1989, {\em Astrophysics of Gaseous Nebulae and Active Galactic Nuclei}, University Science Books



\bibitem []{} Sanders D.B., Mirabel I.F., 1996, ARA\&A, 34, 749

\bibitem[]{} Sanders D.B., Soifer B.T., Elias J.H., Madore B.F., Matthews K., Neugebauer G., Scoville N.Z., 1988a, ApJ, 325, 74

\bibitem[]{} Sanders D.B., Soifer B.T., Elias J.H., Neugebauer G., Matthews K., 1988b, ApJ, 328, L35

\bibitem[]{} Sharpiro S.L., Teukolsky S.A., Black Holes, White Dwarfs and Neutron Stars, Wiley, 1983  

\bibitem[]{} Stark A.A., Gammie C.F., Wilson R.W., Bally J., Linke R.A., Heiles C., Hurwitz M., 1992, ApJS, 79, 77

\bibitem[]{} Surace J.A., Sanders D.B., Vacca W.D., Sylvain V., Mazzarella J.M., 1998, ApJ, 492, 116

\bibitem[]{} Tennant A.F., 1991, NASA Technical Memorandum 4301

\bibitem[]{} Unger S.W., Pedlar A., Axon D.J., Whittle M., Meurs E.J.A., Ward M.J., 1987, MNRAS, 228, 671

\bibitem[]{} van der Kruit P.C., Shostak G.S., 1984, A\&A, 134, 258

\bibitem[]{} Veilleux S., Osterbrock D.E., 1987, ApJS, 63, 295

\bibitem[]{} Veilleux S., 1991, ApJ, 369, 331

\bibitem[]{} Veilleux S., Kim D.-C., Sanders D.B., Mazzarella J.M., Soifer B.T., 1995, ApJS, 98, 171

\bibitem[]{} Veilleux S., Sanders D.B., Kim D.-C., 1997, ApJ, 484, 92

\bibitem[]{} Whittle M., Pedlar A., Meurs E.J.A., Unger S.W., Axon D.J., Ward M.J., 1988, ApJ, 326, 125


\bibitem[]{} Young S., Hough J.H., Axon D.J., Ward M.J., Bailey J.A., 1996b, MNRAS, 280, 291

\bibitem[]{} Young S., Hough J.H., Estathiou A., Wills B.J., Bailey J.A., Ward M.J., Axon D.J., 1996a, MNRAS, 281, 1206 


\end{thebibliography}
\end{document}